\documentclass[final,1p,times]{elsarticle}
\usepackage{amssymb}
\usepackage{amsfonts}
\usepackage{amsmath}
\usepackage{geometry}
\usepackage{bm}
\newtheorem{thm}{Theorem}

\newdefinition{rmk}{Remark}
\newproof{pf}{Proof}
\newproof{pot}{Proof of Theorem}
\begin{document}

\begin{frontmatter}
\title{Lame equation in the algebraic form}

\author{Yoon Seok Choun\corref{cor1}}
\ead{Yoon.Choun@baruch.cuny.edu; ychoun@gradcenter.cuny.edu; ychoun@gmail.com}
\cortext[cor1]{Correspondence to: Baruch College, The City University of New York, Natural Science Department, A506, 17 Lexington Avenue, New York, NY 10010} 
\address{Baruch College, The City University of New York, Natural Science Department, A506, 17 Lexington Avenue, New York, NY 10010}
\begin{abstract}

Lame equation arises from deriving Laplace equation in ellipsoidal coordinates; in other words, it's called ellipsoidal harmonic equation.  Lame functions are applicable to diverse areas such as boundary value problems in ellipsoidal geometry, chaotic Hamiltonian systems, the theory of Bose-Einstein condensates, etc.

In this paper I will apply three term recurrence formula\cite{chou2012b} to the power series expansion in closed forms of Lame function in the algebraic form (infinite series and polynomial) and its integral forms including all higher terms of $A_n$'s. I will show how to transform power series expansion of Lame function to an integral formalism mathematically for cases of infinite series and polynomial.
One interesting observation resulting from the calculations is the fact that a $_2F_1$ function recurs in each of sub-integral forms: the first sub-integral form contains zero term of $A_n's$, the second one contains one term of $A_n$'s, the third one contains two terms of $A_n$'s, etc. Section 6 contains additional examples of application in Lame function. 

This paper is 6th out of 10 in series ``Special functions and three term recurrence formula (3TRF)''. See section 7 for all the papers in the series.  Previous paper in series deals with the power series expansion of Mathieu function and its integral formalism \cite{Chou2012e}. The next paper in the series describes the power series and integral forms of Lame equation in Weierstrass's form and its asymptotic behaviors\cite{Chou2012g}.
\end{abstract}

                                      
\end{frontmatter}                                             
\section{Introduction}
\label{intro}
The sphere is a geometrical perfect shape, the set of points which are all equidistant from its center (a fixed point) in three-dimensional space. In contrast, an ellipsoid is a imperfect one, a surface whose plane sections are all ellipses or circles; the set of points are not same distance from the center of the ellipsoid any more. 
As we all recognize, the nature is nonlinear and imperfect geometrically. For the purpose of simplification, we usually linearize those system in order to take a step to the future with a good numerical approximation. Actually, many geometrical spherical objects (earth, sun, black hole, etc) are not perfectly sphere in nature. The shape of those objects are closely better interpreted by an ellipsoid because of their rotations by themselves. For an example, the ellipsoidal harmonics are represented in calculations of gravitational potential\cite{Romai2001}. However spherical harmonic is preferred over the more mathematically  complex ellipsoid harmonics (the coefficients in a power series expansions of Lame equation have a recursive relation between a 3-term).

In 1837, Gabriel Lame introduced a second ordinary differential equation which has four regular singular points in the method of separation of variables applied to the Laplace equation in elliptic coordinates\cite{Lame1837}. Various authors has called this equation as `Lame equation' or `ellipsoidal harmonic equation'\cite{Erde1955}. 

Previously, there were no analytic solutions in closed forms of Lame equation\cite{Erde1955,Hobs1931,Whit1952}. Using Frobenius  method to obtain analytic solutions (represented either in the algebraic form or Weierstrass's form), the solutions automatically come out a 3-term recurrence relation\cite{Hobs1931,Whit1952}. 
In contrast, most of well-known special functions consist of 2-term recursion relations (Hypergeometric, Bessel, Legendre, Kummer functions, etc).

In this paper we construct the power series expansion of Lame function in closed forms analytically and its integral forms with three-term recurrence formula (3TRF)\cite{chou2012b}. Lame equation is a second-order linear ordinary differential equation of the algebraic form\cite{Lame1837}
\begin{equation}
\frac{d^2{y}}{d{x}^2} + \frac{1}{2}\left(\frac{1}{x-a} +\frac{1}{x-b} + \frac{1}{x-c}\right) \frac{d{y}}{d{x}} +  \frac{-\alpha (\alpha +1) x+q}{4 (x-a)(x-b)(x-c)} y = 0\label{eq:1}
\end{equation}
Lame's equation has four regular singular points: a, b, c and $\infty $. Assume that its solution is
\begin{equation}
y(z)= \sum_{n=0}^{\infty } c_n z^{n+\lambda } \hspace{1cm}\mbox{where}\;\;z=x-a \label{eq:2}
\end{equation}
Plug (\ref{eq:2}) into (\ref{eq:1}).
\begin{equation}
c_{n+1}=A_n \;c_n +B_n \;c_{n-1} \hspace{1cm};n\geq 1\label{eq:3}
\end{equation}
where,
\begin{subequations}
\begin{equation}
A_n = \frac{\frac{1}{4}(\alpha (\alpha +1)a-q)-(2a-b-c)(n+\lambda )^2}{(a-b)(a-c)(n+1+\lambda )(n+\frac{1}{2}+\lambda )}\label{eq:4a}
\end{equation}
\begin{equation}
B_n = \frac{[\alpha -(1-2(n+\lambda ))][\alpha -2(n-1+\lambda )]}{2^2(a-b)(a-c)(n+1+\lambda )(n+\frac{1}{2}+\lambda )}\label{eq:4b}
\end{equation}
\begin{equation}
c_1= A_0 \;c_0\label{eq:4c}
\end{equation}
\end{subequations}
We have two indicial roots which are $\lambda = 0$ and $\frac{1}{2}$. Parameters $b$ and $c$ are identical to each other in (\ref{eq:4a})-(\ref{eq:4c}).
\section{Power series}
\subsection{Polynomial in which makes $B_n$ term terminated}
In this paper I construct the power series expansion, its integral forms and the generating function for the Lame polynomial where $B_n$ term terminated at certain values of index $n$: I treat $q$ as a free variable and $\alpha $ as a fixed value.
\begin{thm}
In Ref.\cite{chou2012b}, the general expression of power series of $y(x)$ for polynomial of $x$ which makes $B_n$ term terminated is
\begin{eqnarray}
 y(x)&=& \sum_{n=0}^{\infty } y_n(x)= y_0(x)+ y_1(x)+ y_2(x)+y_3(x)+\cdots \nonumber\\
&=& c_0 \Bigg\{ \sum_{i_0=0}^{\beta _0} \left( \prod _{i_1=0}^{i_0-1}B_{2i_1+1} \right) x^{2i_0+\lambda } + \sum_{i_0=0}^{\beta _0}\left\{ A_{2i_0} \prod _{i_1=0}^{i_0-1}B_{2i_1+1}  \sum_{i_2=i_0}^{\beta _1} \left( \prod _{i_3=i_0}^{i_2-1}B_{2i_3+2} \right)\right\} x^{2i_2+1+\lambda }\nonumber\\
 && + \sum_{N=2}^{\infty } \Bigg\{ \sum_{i_0=0}^{\beta _0} \Bigg\{A_{2i_0}\prod _{i_1=0}^{i_0-1} B_{2i_1+1} \prod _{k=1}^{N-1} \Bigg( \sum_{i_{2k}= i_{2(k-1)}}^{\beta _k} A_{2i_{2k}+k}\prod _{i_{2k+1}=i_{2(k-1)}}^{i_{2k}-1}B_{2i_{2k+1}+(k+1)}\Bigg)\nonumber\\
 &&\times  \sum_{i_{2N} = i_{2(N-1)}}^{\beta _N} \Bigg( \prod _{i_{2N+1}=i_{2(N-1)}}^{i_{2N}-1} B_{2i_{2N+1}+(N+1)} \Bigg) \Bigg\} \Bigg\} x^{2i_{2N}+N+\lambda }\Bigg\}
  \label{eq:5}
\end{eqnarray}
For a polynomial, we need a condition, which is
\begin{equation}
 B_{2\beta _i + (i+1)}=0 \hspace{1cm} \mathrm{where}\; i,\beta _i =0,1,2,\cdots
 \label{eq:6}
\end{equation}
\end{thm}
In this paper Pochhammer symbol $(x)_n$ is used to represent the rising factorial: $(x)_n = \frac{\Gamma (x+n)}{\Gamma (x)}$.
On above, $ \beta _i$ is an eigenvalue that makes $B_n$ term terminated at certain value of n. (\ref{eq:6}) makes each $y_i(x)$ where $i =0,1,2,\cdots$ as the polynomial in (\ref{eq:5}).

\subsubsection{The case of $\alpha = 2 (2\alpha _i+i +\lambda )$ where $i,\alpha _i =0,1,2,\cdots$}
In (\ref{eq:4a})-(\ref{eq:4c}) replace $\alpha $ by $2 (2\alpha _i+i +\lambda )$. In (\ref{eq:6}) replace index $\beta _i$ by $\alpha _i$. Take the new (\ref{eq:4a})-(\ref{eq:4c}), (\ref{eq:6}) and put them in (\ref{eq:5}) with replacing variable $x$ by $z$.
After the replacement process, the general expression of power series of Lame equation in the algebraic form for polynomial which $B_n$ term is terminated is
\begin{eqnarray}
 y(z)&=& \sum_{n=0}^{\infty } y_n(z)= y_0(z)+ y_1(z)+ y_2(z)+ y_3(z)+\cdots \nonumber\\
&=& c_0 z^{\lambda } \left\{\sum_{i_0=0}^{\alpha _0} \frac{(-\alpha _0)_{i_0} (\alpha _0+ \frac{1}{4}+\lambda )_{i_0}}{(1+\frac{\lambda }{2})_{i_0}(\frac{3}{4} +\frac{\lambda }{2})_{i_0}} \eta ^{i_0} \right.\nonumber\\
&+& \left\{ \sum_{i_0=0}^{\alpha _0} \frac{ (i_0+\frac{\lambda }{2})^2- \Gamma_0^{(P)}}{(i_0+\frac{1}{2}+\frac{\lambda }{2})(i_0+\frac{1}{4}+\frac{\lambda }{2})}\frac{(-\alpha _0)_{i_0} (\alpha _0+\frac{1}{4}+\lambda )_{i_0}}{(1+\frac{\lambda }{2})_{i_0}(\frac{3}{4}+ \frac{\lambda }{2})_{i_0}} \sum_{i_1=i_0}^{\alpha _1} \frac{(-\alpha _1)_{i_1} (\alpha _1+ \frac{5}{4}+\lambda )_{i_1}(\frac{3}{2}+\frac{\lambda}{2})_{i_0}(\frac{5}{4}+\frac{\lambda}{2})_{i_0}}{(-\alpha _1)_{i_0} (\alpha _1+ \frac{5}{4}+\lambda )_{i_0}(\frac{3}{2}+\frac{\lambda}{2})_{i_1}(\frac{5}{4}+\frac{\lambda}{2})_{i_1}} \eta ^{i_1} \right\}\mu  \nonumber\\ 
&+& \sum_{n=2}^{\infty } \left\{ \sum_{i_0=0}^{\alpha _0} \frac{ (i_0+\frac{\lambda }{2})^2- \Gamma_0^{(P)}}{(i_0+\frac{1}{2}+\frac{\lambda }{2})(i_0+\frac{1}{4}+\frac{\lambda }{2})} \frac{(-\alpha _0)_{i_0} (\alpha _0+\frac{1}{4}+\lambda )_{i_0}}{(1+\frac{\lambda }{2})_{i_0}(\frac{3}{4}+ \frac{\lambda }{2})_{i_0}}\right.\nonumber\\
&\times& \prod _{k=1}^{n-1} \left( \sum_{i_k=i_{k-1}}^{\alpha _k} \frac{ (i_k+\frac{k}{2}+ \frac{\lambda }{2})^2- \Gamma_k^{(P)}}{(i_k+\frac{k}{2}+\frac{1}{2}+\frac{\lambda }{2})(i_k+\frac{k}{2}+\frac{1}{4}+\frac{\lambda }{2})}   \frac{(-\alpha _k)_{i_k} (\alpha _k+ k+\frac{1}{4}+\lambda )_{i_k}(1+\frac{k}{2}+\frac{\lambda}{2})_{i_{k-1}}(\frac{3}{4}+\frac{k}{2}+\frac{\lambda}{2})_{i_{k-1}}}{(-\alpha _k)_{i_{k-1}} (\alpha _k+ k+\frac{1}{4}+\lambda )_{i_{k-1}}(1+\frac{k}{2}+\frac{\lambda}{2})_{i_k}(\frac{3}{4}+\frac{k}{2}+\frac{\lambda}{2})_{i_k}}\right) \nonumber\\
&\times& \left.\left. \sum_{i_n= i_{n-1}}^{\alpha _n}\frac{(-\alpha _n)_{i_n} (\alpha _n+ n+\frac{1}{4}+\lambda )_{i_n}(1+\frac{n}{2}+\frac{\lambda}{2})_{i_{n-1}}(\frac{3}{4}+\frac{n}{2}+\frac{\lambda}{2})_{i_{n-1}}}{(-\alpha _n)_{i_{n-1}} (\alpha _n+n+\frac{1}{4}+\lambda )_{i_{n-1}}(1+\frac{n}{2}+\frac{\lambda}{2})_{i_n}(\frac{3}{4}+\frac{n}{2}+\frac{\lambda}{2})_{i_n}} \eta ^{i_n} \right\} \mu ^n \right\}\label{eq:7}
\end{eqnarray}
where
\begin{equation}
\begin{cases} z= x-a \cr
\eta = \frac{-z^2}{(a-b)(a-c)} \cr
\mu  = \frac{-(2a-b-c)z}{(a-b)(a-c)} \cr
\alpha = 2( 2\alpha _i+ i+\lambda )\;\;\mbox{as}\;i,\alpha _i =0,1,2,\cdots \cr
\alpha _i\leq \alpha _j \;\;\mbox{only}\;\mbox{if}\;i\leq j\;\;\mbox{where}\;i,j =0,1,2,\cdots
\end{cases}\nonumber 
\end{equation}
and
\begin{equation}
\begin{cases} 
\Gamma_0^{(P)} = \frac{a}{(2a-b-c)}\left( (\alpha _0+\frac{\lambda }{2})(\alpha _0+\frac{1}{4}+\frac{\lambda }{2})- \frac{q}{2^4 a} \right) \cr
\Gamma_k^{(P)} = \frac{a}{(2a-b-c)}\left( (\alpha _k+\frac{k}{2}+\frac{\lambda }{2})(\alpha _k+\frac{k}{2}+\frac{1}{4}+\frac{\lambda }{2})-\frac{q}{2^4 a} \right)  
\end{cases}\nonumber 
\end{equation}
\subsubsection{The case of $\alpha = -2 (2\alpha _i+i +\lambda )-1$ where $i,\alpha _i =0,1,2,\cdots$}
In (\ref{eq:4a})-(\ref{eq:4c}) replace $\alpha $ by $-2 (2\alpha _i+i +\lambda )-1$. In (\ref{eq:6}) replace  index $\beta _i$ by $\alpha _i$. Take the new (\ref{eq:4a})-(\ref{eq:4c}), (\ref{eq:6}) and put them in (\ref{eq:5}) with replacing variable $x$ by $z$. Its solution is equivalent to (\ref{eq:7}).
Take $c_0$= 1 as $\lambda =0$  for the first independent solution of Lame equation and $\lambda =\frac{1}{2}$ for the second one into (\ref{eq:7}).
\begin{rmk}
The representation in the form of power series expansion of the first kind of independent solution of Lame equation in the algebraic form for the polynomial which makes $B_n$ term terminated  about $x=a$  as $\alpha = 2(2\alpha_j +j) $ or $-2(2\alpha_j +j)-1 $ where $j,\alpha _j =0,1,2,\cdots$ is
\begin{eqnarray}
 y(z)&=& LF_{\alpha _j}\left( a, b, c, q, \alpha = 2(2\alpha_j +j)\; \mbox{or} -2(2\alpha_j +j)-1; z= x-a, \mu = \frac{-(2a-b-c)z}{(a-b)(a-c)}, \eta = \frac{-z^2}{(a-b)(a-c)} \right) \nonumber\\
&=& \sum_{i_0=0}^{\alpha _0} \frac{(-\alpha _0)_{i_0} (\alpha _0+ \frac{1}{4})_{i_0}}{(\frac{3}{4})_{i_0} (1)_{i_0}} \eta ^{i_0} \nonumber\\
&+& \left\{ \sum_{i_0=0}^{\alpha _0} \frac{ i_0^2- \Gamma_0^{(P)}}{(i_0+\frac{1}{2})(i_0+\frac{1}{4})}\frac{(-\alpha _0)_{i_0} (\alpha _0+\frac{1}{4})_{i_0}}{(\frac{3}{4})_{i_0}(1)_{i_0}} \sum_{i_1=i_0}^{\alpha _1} \frac{(-\alpha _1)_{i_1} (\alpha _1+ \frac{5}{4})_{i_1}(\frac{3}{2})_{i_0}(\frac{5}{4})_{i_0}}{(-\alpha _1)_{i_0} (\alpha _1+ \frac{5}{4})_{i_0}(\frac{3}{2})_{i_1}(\frac{5}{4})_{i_1}} \eta ^{i_1} \right\}\mu  \nonumber\\ 
&+& \sum_{n=2}^{\infty } \left\{ \sum_{i_0=0}^{\alpha _0} \frac{ i_0^2- \Gamma_0^{(P)}}{(i_0+\frac{1}{2})(i_0+\frac{1}{4})} \frac{(-\alpha _0)_{i_0} (\alpha _0+\frac{1}{4})_{i_0}}{(\frac{3}{4})_{i_0}(1)_{i_0}}\right.\nonumber\\
&\times& \prod _{k=1}^{n-1} \left( \sum_{i_k=i_{k-1}}^{\alpha _k} \frac{ (i_k+\frac{k}{2})^2- \Gamma_k^{(P)}}{(i_k+\frac{k}{2}+\frac{1}{2})(i_k+\frac{k}{2}+\frac{1}{4})}    \frac{(-\alpha _k)_{i_k} (\alpha _k+ k+\frac{1}{4})_{i_k}(1+\frac{k}{2})_{i_{k-1}}(\frac{3}{4}+\frac{k}{2})_{i_{k-1}}}{(-\alpha _k)_{i_{k-1}} (\alpha _k+ k+\frac{1}{4})_{i_{k-1}}(1+\frac{k}{2})_{i_k}(\frac{3}{4}+\frac{k}{2})_{i_k}}\right) \nonumber\\
&\times& \left. \sum_{i_n= i_{n-1}}^{\alpha _n}\frac{(-\alpha _n)_{i_n} (\alpha _n+ n+\frac{1}{4})_{i_n}(1+\frac{n}{2})_{i_{n-1}}(\frac{3}{4}+\frac{n}{2})_{i_{n-1}}}{(-\alpha _n)_{i_{n-1}} (\alpha _n+n+\frac{1}{4})_{i_{n-1}}(1+\frac{n}{2})_{i_n}(\frac{3}{4}+\frac{n}{2})_{i_n}} \eta ^{i_n} \right\} \mu ^n \nonumber 
\end{eqnarray}
where
\begin{equation}
\begin{cases} 
\Gamma_0^{(P)} = \frac{a}{(2a-b-c)}\left( \alpha _0 (\alpha _0+\frac{1}{4} )- \frac{q}{2^4 a} \right) \cr
\Gamma_k^{(P)} = \frac{a}{(2a-b-c)}\left( (\alpha _k+\frac{k}{2} )(\alpha _k+\frac{k}{2}+\frac{1}{4} )-\frac{q}{2^4 a} \right)  
\end{cases}\nonumber 
\end{equation}
\end{rmk}
For the minimum value of Lame equation of the first kind for a polynomial which makes $B_n$ term terminated about $z=0 $, put $\alpha _0=\alpha _1=\alpha _2=\cdots=0$ in Remark 1.
\begin{eqnarray}
y(z)&=& LF_{0}\left( a, b, c, q, \alpha = 2j\; \mbox{or} -2j-1; z= x-a, \mu = \frac{-(2a-b-c)z}{(a-b)(a-c)}, \eta = \frac{-z^2}{(a-b)(a-c)} \right) \nonumber\\
&=& \sum_{n=0}^{\infty }\frac{\prod_{k=1}^{n}\left( (k-1)^2-\frac{a}{(2a-b-c)} \left( (k-1)\left( k-\frac{1}{2}\right)-\frac{q}{4a}\right)\right)}{\left( \frac{1}{2}\right)_n} \frac{\mu ^n}{n!}\nonumber\\
&=& \; _2F_1\left( \frac{-a-\sqrt{a^2-4(a-b-c)q}}{4(a-b-c)}, \frac{-a+\sqrt{a^2-4(a-b-c)q}}{4(a-b-c)}, \frac{1}{2}, -\frac{(a-b-c)}{(a-b)(a-c)}z\right) \hspace{1cm}\label{ccc:1}\\
&& \mbox{where}\;\;\left|-\frac{(a-b-c)}{(a-b)(a-c)}z\right| < 1 \nonumber
\end{eqnarray} 
For the special case, if $z =- \frac{(a-b)(a-c)}{(a-b-c)}$ and $Re\left( \frac{2a-b-c}{a-b-c}\right)>0 $ in (\ref{ccc:1}),
\begin{eqnarray}
y(z)&=& LF_{0}\left( a, b, c, q, \alpha = 2j\; \mbox{or} -2j-1; z=- \frac{(a-b)(a-c)}{(a-b-c)}, \mu = \frac{(2a-b-c)}{(a-b-c)}, \eta = -\frac{(a-b)(a-c)}{(a-b-c)^2} \right) \nonumber\\
&=& \frac{\sqrt{\pi}\; \Gamma \left( \frac{2a-b-c}{2(a-b-c)}\right)}{\Gamma \left( \frac{1}{2}+\frac{a-\sqrt{a^2-4(a-b-c)q}}{4(a-b-c)}\right) \Gamma \left( \frac{1}{2}+\frac{a+\sqrt{a^2-4(a-b-c)q}}{4(a-b-c)}\right)}   \nonumber
\end{eqnarray}
\begin{rmk}
The representation in the form of power series expansion of the second kind of independent solution of Lame equation in the algebraic form for the polynomial which makes $B_n$ term terminated  about $x=a$ as $\alpha = 2(2\alpha_j +j)+1$ or $-2(2\alpha_j +j+1)$  where $j,\alpha _j =0,1,2,\cdots$ is 
\begin{eqnarray}
y(z)&=& LS_{\alpha _j}\Bigg(  a, b, c, q, \alpha = 2(2\alpha_j +j)+1 \;\mbox{or} -2(2\alpha_j +j+1); z= x-a, \mu = \frac{-(2a-b-c)z}{(a-b)(a-c)}, \eta = \frac{-z^2}{(a-b)(a-c)} \Bigg)\nonumber\\
&=& z^{\frac{1}{2}}  \left\{\sum_{i_0=0}^{\alpha _0} \frac{(-\alpha _0)_{i_0} (\alpha _0+ \frac{3}{4})_{i_0}}{(\frac{5}{4})_{i_0}(1)_{i_0}} \eta ^{i_0} \right. \nonumber\\
&+& \left\{ \sum_{i_0=0}^{\alpha _0} \frac{ (i_0+\frac{1}{4})^2- \Gamma_0^{(P)}}{(i_0+\frac{3}{4})(i_0+\frac{1}{2})}\frac{(-\alpha _0)_{i_0} (\alpha _0+\frac{3}{4})_{i_0}}{(\frac{5}{4})_{i_0}(1)_{i_0}}\sum_{i_1=i_0}^{\alpha _1} \frac{(-\alpha _1)_{i_1} (\alpha _1+ \frac{7}{4})_{i_1}(\frac{7}{4})_{i_0}(\frac{3}{2})_{i_0}}{(-\alpha _1)_{i_0} (\alpha _1+ \frac{7}{4})_{i_0}(\frac{7}{4})_{i_1}(\frac{3}{2})_{i_1}} \eta ^{i_1} \right\}\mu \hspace{2cm} \nonumber\\ 
&+& \sum_{n=2}^{\infty } \left\{ \sum_{i_0=0}^{\alpha _0} \frac{ (i_0+\frac{1}{4})^2- \Gamma_0^{(P)}}{(i_0+\frac{3}{4})(i_0+\frac{1}{2})} \frac{(-\alpha _0)_{i_0} (\alpha _0+\frac{3}{4})_{i_0}}{(\frac{5}{4})_{i_0}(1)_{i_0}}\right.\nonumber\\
&\times& \prod _{k=1}^{n-1} \left( \sum_{i_k=i_{k-1}}^{\alpha _k} \frac{ (i_k+\frac{k}{2}+ \frac{1}{4})^2- \Gamma_k^{(P)}}{(i_k+\frac{k}{2}+\frac{3}{4})(i_k+\frac{k}{2}+\frac{1}{2})} \frac{(-\alpha _k)_{i_k} (\alpha _k+ k+\frac{3}{4})_{i_k}(\frac{5}{4}+\frac{k}{2})_{i_{k-1}}(1+\frac{k}{2})_{i_{k-1}}}{(-\alpha _k)_{i_{k-1}} (\alpha _k+ k+\frac{3}{4})_{i_{k-1}}(\frac{5}{4}+\frac{k}{2})_{i_k}(1+\frac{k}{2})_{i_k}}\right) \nonumber\\
&\times& \left.\left. \sum_{i_n= i_{n-1}}^{\alpha _n}\frac{(-\alpha _n)_{i_n} (\alpha _n+ n+\frac{3}{4})_{i_n}(\frac{5}{4}+\frac{n}{2})_{i_{n-1}}(1+\frac{n}{2})_{i_{n-1}}}{(-\alpha _n)_{i_{n-1}} (\alpha _n+n+\frac{3}{4})_{i_{n-1}}(\frac{5}{4}+\frac{n}{2})_{i_n}(1+\frac{n}{2})_{i_n}} \eta ^{i_n} \right\} \mu ^n \right\}\nonumber 
\end{eqnarray}
where
\begin{equation}
\begin{cases} 
\Gamma_0^{(P)} = \frac{a}{(2a-b-c)}\left( (\alpha _0+\frac{1}{4})(\alpha _0+\frac{1}{2})- \frac{q}{2^4 a} \right) \cr
\Gamma_k^{(P)} = \frac{a}{(2a-b-c)}\left( (\alpha _k+\frac{k}{2}+\frac{1}{4})(\alpha _k+\frac{k}{2}+\frac{1}{2})-\frac{q}{2^4 a} \right)  
\end{cases}\nonumber 
\end{equation}
\end{rmk}
For the minimum value of Lame equation of the second kind for a polynomial which makes $B_n$ term terminated about $z=0 $, put $\alpha _0=\alpha _1=\alpha _2=\cdots=0$ in Remark 2.
\begin{eqnarray}
y(z)&=& LS_{0}\Bigg(  a, b, c, q, \alpha = 2j+1 \;\mbox{or} -2(j+1); z= x-a, \mu = \frac{-(2a-b-c)z}{(a-b)(a-c)}, \eta = \frac{-z^2}{(a-b)(a-c)} \Bigg)\nonumber\\
&=& z^{\frac{1}{2}}\sum_{n=0}^{\infty }\frac{\prod_{k=1}^{n}\left( \left( k-\frac{1}{2}\right)^2-\frac{a}{(2a-b-c)} \left( k\left( k-\frac{1}{2}\right)-\frac{q}{4a}\right)\right)}{\left( \frac{3}{2}\right)_n} \frac{\mu ^n}{n!}\nonumber\\
&=& z^{\frac{1}{2}}\; _2F_1\left( \frac{a-2(b+c)-\sqrt{a^2-4(a-b-c)q}}{4(a-b-c)}, \frac{a-2(b+c)+\sqrt{a^2-4(a-b-c)q}}{4(a-b-c)}, \frac{3}{2}, -\frac{(a-b-c)}{(a-b)(a-c)}z\right) \hspace{1cm}\label{ccc:2}\\
&& \mbox{where}\;\;\left|-\frac{(a-b-c)}{(a-b)(a-c)}z\right| < 1 \nonumber
\end{eqnarray} 
For the special case, if $z =- \frac{(a-b)(a-c)}{(a-b-c)}$ and $Re\left( \frac{2a-b-c}{a-b-c}\right)>0 $ in (\ref{ccc:2}),
\begin{eqnarray}
y(z)&=& LS_{0}\left( a, b, c, q, \alpha = 2j+1 \;\mbox{or} -2(j+1); z=- \frac{(a-b)(a-c)}{(a-b-c)}, \mu = \frac{(2a-b-c)}{(a-b-c)}, \eta = -\frac{(a-b)(a-c)}{(a-b-c)^2} \right) \nonumber\\
&=& \frac{\sqrt{\pi}\; \Gamma \left( \frac{a-2(b+c)}{2(a-b-c)}\right)}{2\; \Gamma \left( \frac{3}{2}- \frac{a-2(b+c)-\sqrt{a^2-4(a-b-c)q}}{4(a-b-c)}\right) \Gamma \left( \frac{3}{2}- \frac{a-2(b+c)+\sqrt{a^2-4(a-b-c)q}}{4(a-b-c)}\right)} \left( - \frac{(a-b)(a-c)}{(a-b-c)}\right)^{\frac{1}{2}}  \nonumber
\end{eqnarray}
(\ref{ccc:1}) and (\ref{ccc:2}) tell us that Lame polynomials in which makes $B_n$ term terminated, for fixed value of $\alpha $, require $\left|-\frac{(a-b-c)}{(a-b)(a-c)}z\right| < 1$ for the convergence of the radius.
\subsection{Infinite series}
\begin{thm}
In Ref.\cite{chou2012b}, the general expression of power series of $y(x)$ for infinite series is
\begin{eqnarray}
y(x)  &=& \sum_{n=0}^{\infty } y_{n}(x)= y_0(x)+ y_1(x)+ y_2(x)+ y_3(x)+\cdots \nonumber\\
&=& c_0 \Bigg\{ \sum_{i_0=0}^{\infty } \left( \prod _{i_1=0}^{i_0-1}B_{2i_1+1} \right) x^{2i_0+\lambda } 
+ \sum_{i_0=0}^{\infty }\left\{ A_{2i_0} \prod _{i_1=0}^{i_0-1}B_{2i_1+1}  \sum_{i_2=i_0}^{\infty } \left( \prod _{i_3=i_0}^{i_2-1}B_{2i_3+2} \right)\right\} x^{2i_2+1+\lambda }  \nonumber\\
&& + \sum_{N=2}^{\infty } \Bigg\{ \sum_{i_0=0}^{\infty } \Bigg\{A_{2i_0}\prod _{i_1=0}^{i_0-1} B_{2i_1+1} 
 \prod _{k=1}^{N-1} \Bigg( \sum_{i_{2k}= i_{2(k-1)}}^{\infty } A_{2i_{2k}+k}\prod _{i_{2k+1}=i_{2(k-1)}}^{i_{2k}-1}B_{2i_{2k+1}+(k+1)}\Bigg)\nonumber\\
&& \times  \sum_{i_{2N} = i_{2(N-1)}}^{\infty } \Bigg( \prod _{i_{2N+1}=i_{2(N-1)}}^{i_{2N}-1} B_{2i_{2N+1}+(N+1)} \Bigg) \Bigg\} \Bigg\} x^{2i_{2N}+N+\lambda }\Bigg\} 
\label{eq:12}
\end{eqnarray}
\end{thm}
In (\ref{eq:12}) replace a variable $x$ by $z$ and substitute (\ref{eq:4a})-(\ref{eq:4c}) into new (\ref{eq:12}). 
The general expression of power series of Lame equation in the algebraic form for infinite series is given by
\begin{eqnarray}
 y(z)&=& \sum_{n=0}^{\infty } y_n(z)= y_0(z)+ y_1(z)+ y_2(z)+ y_3(z)+\cdots\nonumber\\
&=& c_0 z^{\lambda } \left\{\sum_{i_0=0}^{\infty } \frac{(-\frac{\alpha }{4}+\frac{\lambda }{2})_{i_0} (\frac{\alpha }{4}+\frac{1}{4}+\frac{\lambda }{2})_{i_0}}{(1+\frac{\lambda }{2})_{i_0}(\frac{3}{4}+ \frac{\lambda }{2})_{i_0}} \eta^{i_0} \right.\nonumber\\
&+& \left\{\sum_{i_0=0}^{\infty } \frac{ (i_0 +\frac{\lambda }{2})^2 -\Gamma ^{(I)}}{(i_0+ \frac{1}{2}+ \frac{\lambda }{2})(i_0 + \frac{1}{4}+ \frac{\lambda }{2})}  \frac{(-\frac{\alpha }{4}+\frac{\lambda }{2})_{i_0} (\frac{\alpha }{4}+\frac{1}{4}+\frac{\lambda }{2})_{i_0}}{(1+\frac{\lambda }{2})_{i_0}(\frac{3}{4} +\frac{\lambda }{2})_{i_0}} \sum_{i_1=i_0}^{\infty } \frac{(-\frac{\alpha }{4} + \frac{1}{2} + \frac{\lambda }{2})_{i_1}(\frac{\alpha }{4}+\frac{3}{4}+ \frac{\lambda }{2})_{i_1}(\frac{3}{2}+\frac{\lambda }{2})_{i_0}(\frac{5}{4}+ \frac{\lambda }{2})_{i_0}}{(-\frac{\alpha }{4} + \frac{1}{2} + \frac{\lambda }{2})_{i_0}(\frac{\alpha }{4}+\frac{3}{4}+ \frac{\lambda }{2})_{i_0}(\frac{3}{2}+\frac{\lambda }{2})_{i_1}(\frac{5}{4}+ \frac{\lambda }{2})_{i_1}} \eta ^{i_1} \right\}\mu \nonumber\\
&+& \sum_{n=2}^{\infty } \left\{ \sum_{i_0=0}^{\infty } \frac{ (i_0+\frac{\lambda }{2})^2-\Gamma ^{(I)}}{(i_0+\frac{1}{2}+\frac{\lambda }{2})(i_0+\frac{1}{4}+\frac{\lambda }{2})} \frac{(-\frac{\alpha }{4}+\frac{\lambda }{2})_{i_0} (\frac{\alpha }{4}+\frac{1}{4}+\frac{\lambda }{2})_{i_0}}{(1+\frac{\lambda }{2})_{i_0}(\frac{3}{4} +\frac{\lambda }{2})_{i_0}} \right.\nonumber\\
&\times& \prod _{k=1}^{n-1} \left( \sum_{i_k=i_{k-1}}^{\infty } \frac{ (i_k+\frac{k}{2}+ \frac{\lambda }{2})^2- \Gamma ^{(I)}}{(i_k+\frac{k}{2}+\frac{1}{2}+\frac{\lambda }{2})(i_k+\frac{k}{2}+\frac{1}{4}+\frac{\lambda }{2})}    \frac{(-\frac{\alpha }{4}+\frac{k}{2}+\frac{\lambda }{2})_{i_k} (\frac{\alpha }{4}+\frac{k}{2}+\frac{1}{4}+\frac{\lambda }{2})_{i_k}(1+\frac{k}{2}+\frac{\lambda}{2})_{i_{k-1}}(\frac{k}{2}+\frac{3}{4}+\frac{\lambda}{2})_{i_{k-1}}}{(-\frac{\alpha }{4}+\frac{k}{2}+\frac{\lambda }{2})_{i_{k-1}} (\frac{\alpha }{4}+\frac{k}{2}+\frac{1}{4}+\frac{\lambda }{2})_{i_{k-1}}(1+\frac{k}{2}+\frac{\lambda}{2})_{i_k}(\frac{k}{2}+\frac{3}{4}+\frac{\lambda}{2})_{i_k}}\right) \nonumber\\
&\times& \left.\left. \sum_{i_n= i_{n-1}}^{\infty }\frac{(-\frac{\alpha }{4}+\frac{n}{2}+\frac{\lambda }{2})_{i_n} (\frac{\alpha }{4}+\frac{n}{2}+\frac{1}{4}+\frac{\lambda }{2})_{i_n}(1+\frac{n}{2}+\frac{\lambda}{2})_{i_{n-1}}(\frac{n}{2}+\frac{3}{4}+\frac{\lambda}{2})_{i_{n-1}}}{(-\frac{\alpha }{4}+\frac{n}{2}+\frac{\lambda }{2})_{i_{n-1}} (\frac{\alpha }{4}+\frac{n}{2}+\frac{1}{4}+\frac{\lambda }{2})_{i_{n-1}}(1+\frac{n}{2}+\frac{\lambda}{2})_{i_n}(\frac{n}{2}+\frac{3}{4}+\frac{\lambda}{2})_{i_n}} \eta ^{i_n} \right\} \mu ^n \right\}\label{eq:13}
\end{eqnarray}
where
\begin{equation}
\begin{cases} z= x-a \cr
\eta = \frac{-z^2}{(a-b)(a-c)} \cr
\mu  = \frac{-(2a-b-c)z}{(a-b)(a-c)} \cr
\Gamma ^{(I)} = \frac{a}{2^4(2a-b-c)}\left( \alpha (\alpha +1) -\frac{q}{a}\right) 
\end{cases}\nonumber  
\end{equation}
Take $c_0$= 1 as $\lambda =0$  for the first independent solution of Lame equation and $\lambda =\frac{1}{2}$ for the second one into (\ref{eq:13}). 
\begin{rmk}
The representation in the form of power series expansion of the first kind of independent solution of Lame equation in the algebraic form for the infinite series about $x=a$ is
\begin{eqnarray}
 y(z)&=& LF \left( a, b, c, q, \alpha, \Gamma ^{(I)} = \frac{a}{2^4(2a-b-c)}\left( \alpha (\alpha +1) -\frac{q}{a}\right); z=x-a, \mu = \frac{-(2a-b-c)z}{(a-b)(a-c)}, \eta = \frac{-z^2}{(a-b)(a-c)} \right) \nonumber\\
&=& \sum_{i_0=0}^{\infty } \frac{(-\frac{\alpha }{4})_{i_0} (\frac{\alpha }{4}+\frac{1}{4})_{i_0}}{(\frac{3}{4})_{i_0}(1)_{i_0}} \eta^{i_0} \nonumber\\
&+& \left\{\sum_{i_0=0}^{\infty } \frac{ i_0^2 - \Gamma ^{(I)}}{(i_0+ \frac{1}{2})(i_0 + \frac{1}{4})}\frac{(-\frac{\alpha }{4})_{i_0} (\frac{\alpha }{4}+\frac{1}{4})_{i_0}}{(\frac{3}{4})_{i_0}(1)_{i_0}} \sum_{i_1=i_0}^{\infty } \frac{(-\frac{\alpha }{4} + \frac{1}{2})_{i_1}(\frac{\alpha }{4}+\frac{3}{4})_{i_1}(\frac{3}{2})_{i_0}(\frac{5}{4})_{i_0}}{(-\frac{\alpha }{4} + \frac{1}{2})_{i_0}(\frac{\alpha }{4}+\frac{3}{4})_{i_0}(\frac{3}{2})_{i_1}(\frac{5}{4})_{i_1}} \eta ^{i_1} \right\} \mu \nonumber\\
&+& \sum_{n=2}^{\infty } \left\{ \sum_{i_0=0}^{\infty } \frac{ i_0^2- \Gamma ^{(I)}}{(i_0+\frac{1}{2})(i_0+\frac{1}{4})} \frac{(-\frac{\alpha }{4})_{i_0} (\frac{\alpha }{4}+\frac{1}{4})_{i_0}}{(\frac{3}{4})_{i_0}(1)_{i_0}} \right.\nonumber\\
&\times& \prod _{k=1}^{n-1} \left( \sum_{i_k=i_{k-1}}^{\infty } \frac{ (i_k+\frac{k}{2})^2- \Gamma ^{(I)}}{(i_k+\frac{k}{2}+\frac{1}{2})(i_k+\frac{k}{2}+\frac{1}{4})} \frac{(-\frac{\alpha }{4}+\frac{k}{2})_{i_k} (\frac{\alpha }{4}+\frac{1}{4}+\frac{k}{2})_{i_k}(1+\frac{k}{2})_{i_{k-1}}(\frac{3}{4}+\frac{k}{2})_{i_{k-1}}}{(-\frac{\alpha }{4}+\frac{k}{2})_{i_{k-1}} (\frac{\alpha }{4}+\frac{1}{4}+\frac{k}{2})_{i_{k-1}}(1+\frac{k}{2})_{i_k}(\frac{3}{4}+\frac{k}{2})_{i_k}}\right) \nonumber\\ 
&\times& \left. \sum_{i_n= i_{n-1}}^{\infty }\frac{(-\frac{\alpha }{4}+\frac{n}{2})_{i_n} (\frac{\alpha }{4}+\frac{1}{4}+\frac{n}{2})_{i_n}(1+\frac{n}{2})_{i_{n-1}}(\frac{3}{4}+\frac{n}{2})_{i_{n-1}}}{(-\frac{\alpha }{4}+\frac{n}{2})_{i_{n-1}} (\frac{\alpha }{4}+\frac{1}{4}+\frac{n}{2})_{i_{n-1}}(1+\frac{n}{2})_{i_n}(\frac{3}{4}+\frac{n}{2})_{i_n}} \eta ^{i_n} \right\} \mu ^n \nonumber
\end{eqnarray}
\end{rmk}
\begin{rmk}
The representation in the form of power series expansion of the second kind of independent solution of Lame equation in the algebraic form for the infinite series about $x=a$ is
\begin{eqnarray}
y(z)&=&  LS \left( a, b, c, q, \alpha, \Gamma ^{(I)} = \frac{a}{2^4(2a-b-c)}\left( \alpha (\alpha +1) -\frac{q}{a}\right); z=x-a, \mu = \frac{-(2a-b-c)z}{(a-b)(a-c)}, \eta = \frac{-z^2}{(a-b)(a-c)} \right) \nonumber\\
&=& z^{\frac{1}{2}} \left\{\sum_{i_0=0}^{\infty } \frac{(-\frac{\alpha }{4}+\frac{1}{4})_{i_0} (\frac{\alpha }{4}+\frac{1}{2})_{i_0}}{(\frac{5 }{4})_{i_0}(1)_{i_0}} \eta^{i_0} \right. \nonumber\\
&+& \left\{\sum_{i_0=0}^{\infty } \frac{ (i_0 +\frac{1}{4})^2 - \Gamma ^{(I)}}{(i_0+ \frac{3}{4})(i_0 + \frac{1}{2})}\frac{(-\frac{\alpha }{4}+\frac{1}{4})_{i_0} (\frac{\alpha }{4}+\frac{1}{2})_{i_0}}{(\frac{5}{4})_{i_0}(1)_{i_0}} \sum_{i_1=i_0}^{\infty } \frac{(-\frac{\alpha }{4} + \frac{3}{4})_{i_1}(\frac{\alpha }{4}+1)_{i_1}(\frac{7}{4})_{i_0}(\frac{3}{2})_{i_0}}{(-\frac{\alpha }{4} + \frac{3}{4})_{i_0}(\frac{\alpha }{4}+1)_{i_0}(\frac{7}{4})_{i_1}(\frac{3}{2})_{i_1}} \eta ^{i_1} \right\} \mu \nonumber\\
&+& \sum_{n=2}^{\infty } \Bigg\{ \sum_{i_0=0}^{\infty } \frac{ (i_0+\frac{1}{4})^2- \Gamma ^{(I)}}{(i_0+\frac{3}{4})(i_0+\frac{1}{2})} \frac{(-\frac{\alpha }{4}+\frac{1}{4})_{i_0} (\frac{\alpha }{4}+\frac{1}{2})_{i_0}}{(\frac{5}{4})_{i_0}(1)_{i_0}} \nonumber\\
&\times& \prod _{k=1}^{n-1} \left( \sum_{i_k=i_{k-1}}^{\infty } \frac{ (i_k+\frac{k}{2}+ \frac{1}{4})^2- \Gamma ^{(I)}}{(i_k+\frac{k}{2}+\frac{3}{4})(i_k+\frac{k}{2}+\frac{1}{2})}  \frac{(-\frac{\alpha }{4}+\frac{k}{2}+\frac{1}{4})_{i_k} (\frac{\alpha }{4}+\frac{k}{2}+\frac{1}{2})_{i_k}(\frac{5}{4}+\frac{k}{2})_{i_{k-1}}(1+\frac{k}{2})_{i_{k-1}}}{(-\frac{\alpha }{4}+\frac{k}{2}+\frac{1}{4})_{i_{k-1}} (\frac{\alpha }{4}+\frac{k}{2}+\frac{1}{2})_{i_{k-1}}(\frac{5}{4}+\frac{k}{2})_{i_k}(1+\frac{k}{2})_{i_k}}\right) \nonumber\\
&\times& \left. \left. \sum_{i_n= i_{n-1}}^{\infty }\frac{(-\frac{\alpha }{4}+\frac{n}{2}+\frac{1}{4})_{i_n} (\frac{\alpha }{4}+\frac{n}{2}+\frac{1}{2})_{i_n}(\frac{5}{4}+\frac{n}{2})_{i_{n-1}}(1+\frac{n}{2})_{i_{n-1}}}{(-\frac{\alpha }{4}+\frac{n}{2}+\frac{1}{4})_{i_{n-1}} (\frac{\alpha }{4}+\frac{n}{2}+\frac{1}{2})_{i_{n-1}}(\frac{5}{4}+\frac{n}{2})_{i_n}(1+\frac{n}{2})_{i_n}} \eta ^{i_n} \right\} \mu ^n \right\}\nonumber
\end{eqnarray}
\end{rmk}
\section{Asymptotic behavior of the function $y(z=x-a)$ and the boundary condition for $x$}
Now let's test for convergence of an infinite series of the analytic function $y(z)$.\footnote {For asymptotic expansions in closed forms of the multi-term recurrence relation in a linear ODE, its analytic solution is available in chapter 3 of Ref.\cite{Choun2014}.} As $n\gg 1$ (for sufficiently large, like an index $n$ is closed to infinity, or treat as $n\rightarrow \infty $), (\ref{eq:3})--(\ref{eq:4b}) are asymptotically equal to
\begin{subequations}
\begin{equation}
c_{n+1}=A\;c_n +B\;c_{n-1} \hspace{1cm};n\geq 1 \label{er:100}
\end{equation}
where
\begin{equation}
\lim_{n\gg 1} A_n = A= \frac{-(2a-b-c)}{(a-b)(a-c)} \hspace{2cm} \lim_{n\gg 1} B_n = B= \frac{-1}{(a-b)(a-c)}\label{eq:38}
\end{equation}
\end{subequations}
Substitute (\ref{eq:38}) into (\ref{er:100}) by letting $c_1\sim  A\;c_0$.\footnote{We only have the sense of curiosity about an  asymptotic series as  $n\gg 1$ for given $z$. Actually, $c_1 =  A_0c_0$. But for a huge value of an index $n$, we treat the coefficient $c_1$ as $ Ac_0$ for simple computations.} For $n=0,1,2,\cdots$, it gives
\begin{equation}
\begin{tabular}{ l }
  \vspace{2 mm}
  $c_0$ \\
  \vspace{2 mm}
  $c_1 = A c_0 $ \\
  \vspace{2 mm}
  $c_2 = (A^2+ B)c_0 $ \\
  \vspace{2 mm}
  $c_3 = (A^3 + 2AB)c_0 $ \\
  \vspace{2 mm}
  $c_4 = (A^4 + 3 A^2B + B^2)c_0 $ \\
  \vspace{2 mm}
  $c_5 = (A^5 + 4A^3B + 3AB^2)c_0 $ \\
  \vspace{2 mm}
  $c_6 = (A^6+ 5A^4B+ 6A^2B^2+ B^3) c_0 $ \\
  \vspace{2 mm}
  $c_7 = (A^7+ 6A^5B + 10A^3B^2+ 4AB^3)c_0 $ \\
  \vspace{2 mm}
  $c_8 = (A^8+ 7A^6B+ 15A^4B^2+ 10A^2B^3+ B^4)c_0 $ \\
  \vspace{2 mm}                         
 \hspace{2 mm} \vdots \hspace{3cm} \vdots \\
\end{tabular}\label{eq:39}
\end{equation}
If a series solution of a linear differential equation is absolutely convergent, we can rearrange of its terms for the series solution. Indeed, the sum of any arbitrary series is equivalent to the sum of the initial series.

With reminding the above mathematical phenomenon, let assume that a series solution of Lame equation is absolutely convergent.
The sequences $c_n$ consists of combinations $A$ and $B$ in (\ref{eq:39}). First observe the term inside
parentheses of sequence $c_n$ which does not include any $A_n$'s: $c_n$ with even index ($c_0$,$c_2$,
$c_4$,$\cdots$). 

\begin{equation}
\begin{tabular}{  l  }
  \vspace{2 mm}
  $c_0$ \\
  \vspace{2 mm}
  $c_2 = B c_0  $ \\
  \vspace{2 mm}
  $c_4 = B^2 c_0  $ \\
  \vspace{2 mm}
  $c_6 = B^3c_0 $ \\
  \vspace{2 mm}
  $c_8 = B^4c_0 $\\
  \vspace{2 mm}
  $c_{10} = B^5c_0 $ \\
  \hspace{2 mm}
  \large{\vdots}\hspace{1cm}\large{\vdots} \\ 
\end{tabular}\label{eq:40}
\end{equation}
When a function $y(z)$, analytic at z=0, is expanded in a power series z=0, we write
\begin{equation}
y(z)= \sum_{m=0}^{\infty } y_m(z) \label{eq:41}
\end{equation}
where
\begin{equation}
y_m(z)= \sum_{n=0}^{\infty } c_n^m z^{n}\label{eq:42}
\end{equation}
Put(\ref{eq:40}) in (\ref{eq:42}) putting $m=0$. 
\begin{equation}
y_0(z)= c_0 \sum_{n=0}^{\infty } \left(Bz^2\right)^n \label{eq:43}
\end{equation}
Observe the terms inside parentheses of sequence $c_n$ which include one term of $A_n$'s in (\ref{eq:39}): $c_n$ with odd index ($c_1$, $c_3$, $c_5$,$\cdots$). 

\begin{equation}
\begin{tabular}{  l  }
  \vspace{2 mm}
  $c_1= A c_0$ \\
  \vspace{2 mm}
  $c_3 = 2AB c_0  $ \\
  \vspace{2 mm}
  $c_5 = 3AB^2c_0  $ \\
  \vspace{2 mm}
  $c_7 = 4AB^3c_0  $ \\
  \vspace{2 mm}
  $c_9 = 5AB^4c_0 $\\
  \hspace{2 mm}
  \large{\vdots}\hspace{1cm}\large{\vdots} \\ 
\end{tabular}\label{eq:44}
\end{equation}
Put (\ref{eq:44}) in (\ref{eq:42}) putting $m=1$.
\begin{equation}
y_1(z)= c_0 A z\sum_{n=0}^{\infty } \frac{(n+1)}{1!} \left(Bz^2\right)^n \label{eq:45}
\end{equation}
Observe the terms inside parentheses of sequence $c_n$ which include two terms of $A_n$'s in (\ref{eq:39}): $c_n$ with even index ($c_2$, $c_4$, $c_6$,$\cdots$).  

\begin{equation}
\begin{tabular}{  l  }
  \vspace{2 mm}
  $c_2= A^2 c_0$ \\
  \vspace{2 mm}
  $c_4 = 3A^2B c_0  $ \\
  \vspace{2 mm}
  $c_6 = 6A^2B^2c_0  $ \\
  \vspace{2 mm}
  $c_8 = 10A^2B^3c_0  $ \\
  \vspace{2 mm}
  $c_{10} = 15A^2B^4c_0 $\\
  \vspace{2 mm}
  \large{\vdots}\hspace{1cm}\large{\vdots} \\ 
\end{tabular}\label{eq:46}
\end{equation}
Put (\ref{eq:46}) in (\ref{eq:42}) putting $m=2$.
\begin{equation}
y_2(z)= c_0 \left(Az\right)^2\sum_{n=0}^{\infty } \frac{(n+1)(n+2)}{2!} \left(Bz^2\right)^n \label{eq:47}
\end{equation}
Similarly, the asymptotic function $y_3(z)$ for three terms of $A$'s is given by
\begin{equation}
y_3(z)= c_0 \left( Az \right)^3 \sum_{n=0}^{\infty } \frac{(n+1)(n+2)(n+3)}{3!}\left( Bz^2 \right)^n
\label{et:14}
\end{equation} 
By repeating this process for all higher terms of $A$'s, we obtain every $y_m(x)$ terms where $m \geq 3$. Substitute (\ref{eq:43}), (\ref{eq:45}), (\ref{eq:47}), (\ref{et:14}) and including all $y_m(x)$ terms where $m \geq 3$ into (\ref{eq:41}).
\begin{eqnarray}
y(z)&=& \sum_{n=0}^{\infty } c_n z^{n}= y_0(z)+ y_1(z)+ y_2(z)+y_3(z)+\cdots \nonumber\\
&=& \sum_{n=0}^{\infty } \sum_{m=0}^{\infty } \frac{(n+m)!}{n!\;m!} \tilde{x}^n \tilde{y}^m \hspace{.6cm} \mbox{where}\;c_0=1, \tilde{x}= Bz^2 \;\mbox{and} \; \tilde{y}= Az \label{eq:48}
\end{eqnarray}
By definition, a real or complex series $\sum_{n=0}^{\infty } u_n$ is said to converge absolutely if the series of muduli $\sum_{n=0}^{\infty } |u_n|$  converge. And the series of absolute values (\ref{eq:48}) is
\begin{equation}
\sum_{n=0}^{\infty } \sum_{m=0}^{\infty } \frac{(n+m)!}{n!\;m!} |\tilde{x}|^n |\tilde{y}|^m = \sum_{r=0}^{\infty } (|\tilde{x}| + |\tilde{y}|)^r  \nonumber
\end{equation} 
This double series is absolutely convergent for $|\tilde{x}| + |\tilde{y}| <1$.
Substitute (\ref{eq:38}) in (\ref{eq:48}) with $z=x-a$.
\begin{equation}
\lim_{n\gg 1}y(z)= \frac{1}{1+\left(\frac{(x-a)^2}{(a-b)(a-c)}+ \frac{(2a-b-c)(x-a)}{(a-b)(a-c)}\right)}  \label{eq:49}
\end{equation}
(\ref{eq:49}) is geometric series. Its condition of convergence of it is
\begin{equation}
\left| \frac{(x-a)^2}{(a-b)(a-c)}\right| + \left| \frac{(2a-b-c)(x-a)}{(a-b)(a-c)}\right| <1 \label{eq:50}
\end{equation}
The coefficients $a$, $b$ and $c$ decide the range of independent variable $x$ as we see (\ref{eq:50}). More precisely, 
\begin{table}[h]
\begin{center}
\tabcolsep 5.8pt
\footnotesize
\begin{tabular}{|l*{1}{c}|r|}
\hline
Range of the coefficients $a$, $b$ and $c$ & Range of the independent variable $z= x-a$ \\
\hline \hline
As $a=b$ or $a=c$ & no solution \\
\hline 
& \\
As $a-b>0$, $a-c>0$  & $|z|<\frac{-(2a-b-c)+\sqrt{(2a-b-c)^2+4(a-b)(a-c)}}{2}$  \\
\hline 
& \\
As $a-b<0$, $a-c<0$ & $|z|<\frac{(2a-b-c)+\sqrt{(2a-b-c)^2+4(a-b)(a-c)}}{2}$ \\
\hline 
& \\
As $a-b<0$, $a-c>0$, $2a-b-c>0$ & $|z|<-(a-b)$ \\
\hline
& \\ 
As $a-b>0$, $a-c<0$, $2a-b-c<0$ & $|z|<(a-b)$ \\
\hline
& \\
As $b=c>0$ & $|z|<(-1+\sqrt{2})(a-b)$ \\
\hline
& \\
As $b=c<0$ & $|z|<(1-\sqrt{2})(a-b)$ \\
\hline
& \\
\hline
\end{tabular}
\end{center}
\caption{Boundary condition of $x$ for an infinite series of Lame equation in the algebraic form about $x=a$}
\end{table}

In the case of $2a-b-c=0$, (\ref{eq:49}) turns to be
\begin{equation}
\lim_{n\gg 1}y(z)= \frac{1}{1+ \frac{(x-a)^2}{(a-b)(a-c)} }  \nonumber
\end{equation}
where $\left| \frac{(x-a)^2}{(a-b)(a-c)} \right|<1$ with $a \ne b $ and $a \ne c$.

If $|a-b|\approx M$ ($M$ is sufficiently large positive value) and $|a-c|\ll 1$ (for extremely small value, like $|a-c|$ is closed to zero), (\ref{eq:49}) is approximately equal to
\begin{equation}
\lim_{n\gg 1}y(z)\approx  \frac{1}{1+ \frac{(2a-b-c)(x-a)}{(a-b)(a-c)} }  \nonumber
\end{equation}
where $\left| \frac{(2a-b-c)(x-a)}{(a-b)(a-c)} \right|<1$ with $a \ne b $ and $a \ne c$.
\section{Integral Formalism}
\subsection{Polynomial in which makes $B_n$ term terminated}
Now let's investigate the integral formalism for the polynomial case of $B_n$ term terminated at certain eigenvalue. There is a generalized hypergeometric function which is:  In this paper Pochhammer symbol $(x)_n$ is used to represent the rising factorial: $(x)_n = \frac{\Gamma (x+n)}{\Gamma (x)}$.
\begin{eqnarray}
I_l &=& \sum_{i_l= i_{l-1}}^{\alpha _l} \frac{(-\alpha _l)_{i_l}(\alpha _l+l+\frac{1}{4}+\lambda )_{i_l}(1+\frac{l}{2}+\frac{\lambda }{2})_{i_{l-1}}(\frac{3}{4}+\frac{l}{2} +\frac{\lambda }{2})_{i_{l-1}}}{(-\alpha _l)_{i_{l-1}}(\alpha _l+l+\frac{1}{4}+\lambda )_{i_{l-1}}(1+\frac{l}{2}+\frac{\lambda }{2})_{i_l}(\frac{3}{4}+\frac{l}{2} +\frac{\lambda }{2})_{i_l}} \eta^{i_l} \label{eq:63}\\
&=& \eta^{i_{l-1}} 
\sum_{j=0}^{\infty } \frac{B(i_{l-1}+\frac{l}{2}-\frac{1}{4}+\frac{\lambda }{2},j+1) B(i_{l-1}+\frac{l}{2}+\frac{\lambda }{2},j+1)(i_{l-1}-\alpha _l)_j (\alpha _l+i_{l-1}+l+\frac{l}{4}+\lambda )_j}{(i_{l-1}+\frac{l}{2}-\frac{1}{4}+\frac{\lambda }{2})^{-1}(i_{l-1}+\frac{l}{2}+ \frac{\lambda }{2})^{-1}(1)_j \;j!} \eta^j \nonumber
\end{eqnarray}
By using integral form of beta function,
\begin{subequations}
\begin{equation}
B\left(i_{l-1}+\frac{l}{2}-\frac{1}{4}+\frac{\lambda }{2},j+1\right)= \int_{0}^{1} dt_l\;t_l^{i_{l-1}+\frac{l}{2}-\frac{5}{4}+\frac{\lambda }{2}} (1-t_l)^j\label{eq:64a}
\end{equation}
\begin{equation}
B\left(i_{l-1}+\frac{l}{2} +\frac{\lambda }{2},j+1\right)= \int_{0}^{1} du_l\;u_l^{i_{l-1}+\frac{l}{2}-1+\frac{\lambda }{2}} (1-u_l)^j\label{eq:64b}
\end{equation}
\end{subequations}
Substitute (\ref{eq:64a}) and (\ref{eq:64b}) into (\ref{eq:63}). And divide $(i_{l-1}+\frac{l}{2}-\frac{1}{4}+\frac{\lambda }{2})(i_{l-1}+\frac{l}{2}+ \frac{\lambda }{2})$ into $I_l$.
\begin{eqnarray}
K_l &=& \frac{1}{(i_{l-1}+\frac{l}{2}-\frac{1}{4}+\frac{\lambda }{2})(i_{l-1}+\frac{l}{2}+ \frac{\lambda }{2})}\nonumber\\
&&\times \sum_{i_l= i_{l-1}}^{\alpha _l} \frac{(-\alpha _l)_{i_l}(\alpha _l+l+\frac{1}{4}+\lambda )_{i_l}(1+\frac{l}{2}+\frac{\lambda }{2})_{i_{l-1}}(\frac{3}{4}+\frac{l}{2} +\frac{\lambda }{2})_{i_{l-1}}}{(-\alpha _l)_{i_{l-1}}(\alpha _l+l+\frac{1}{4}+\lambda )_{i_{l-1}}(1+\frac{l}{2}+\frac{\lambda }{2})_{i_l}(\frac{3}{4}+\frac{l}{2} +\frac{\lambda }{2})_{i_l}} \eta^{i_l}\nonumber\\
&=&  \int_{0}^{1} dt_l\;t_l^{\frac{l}{2}-\frac{5}{4}+\frac{\lambda }{2}} \int_{0}^{1} du_l\;u_l^{\frac{l}{2}-1+\frac{\lambda }{2}} (t_l u_l\eta )^{i_{l-1}}\nonumber\\
&&\times \sum_{j=0}^{\infty } \frac{(i_{l-1}-\alpha _l)_j (i_{l-1}+l+\alpha _l+\frac{1}{4}+\lambda )_j}{(1)_j \;j!} [\eta (1-t_l)(1-u_l)]^j
\label{eq:65}
\end{eqnarray}
The integral form of hypergeometric function is
\begin{eqnarray}
_2F_1 \left( \alpha ,\beta ; \gamma ; z \right) &=& \sum_{n=0}^{\infty } \frac{(\alpha )_n (\beta )_n}{(\gamma )_n (n!)} z^n \label{eq:66}\\
&=& -\frac{1}{2\pi i} \frac{\Gamma(1-\alpha ) \Gamma(\gamma )}{\Gamma (\gamma -\alpha )} \oint dv_l\;(-v_l)^{\alpha -1} (1-v_l)^{\gamma -\alpha -1} (1-zv_l)^{-\beta }\nonumber\\
&& \mbox{where} \;\mbox{Re}(\gamma -\alpha )>0 \nonumber
\end{eqnarray}
replaced $\alpha $, $\beta $, $\gamma $ and z by $i_{l-1}-\alpha _l$, $i_{l-1}+l+\alpha _l+\frac{l}{4}+\lambda $, 1 and $\eta (1-t_l)(1-u_l)$ in (\ref{eq:66})
\begin{eqnarray}
&& \sum_{j=0}^{\infty } \frac{(i_{l-1}-\alpha _l)_j (i_{l-1}+l+\alpha _l+\frac{l}{4}+\lambda )_j}{(1)_j \;j!} [\eta (1-t_l)(1-u_l)]^j \nonumber\\
&=& \frac{1}{2\pi i} \oint dv_l\;\frac{1}{v_l} (1-\eta v_l (1-t_l)(1-u_l))^{-(l+\frac{1}{4}+\lambda )} \left(\frac{1- \frac{1}{v_l}}{1-\eta v_l (1-t_l)(1-u_l)}\right)^{\alpha_l} \nonumber\\
&&\times \left(\frac{1}{(1- \frac{1}{v_l})(1-\eta v_l (1-t_l)(1-u_l))}\right)^{i_{l-1}} \label{eq:67}
\end{eqnarray}
Substitute (\ref{eq:67}) into (\ref{eq:65}).
\begin{eqnarray}
K_l &=& \frac{1}{(i_{l-1}+\frac{l}{2}-\frac{1}{4}+\frac{\lambda }{2})(i_{l-1}+\frac{l}{2}+ \frac{\lambda }{2})} \sum_{i_l= i_{l-1}}^{\alpha _l} \frac{(-\alpha _l)_{i_l}(\alpha _l+l+\frac{1}{4}+\lambda )_{i_l}(1+\frac{l}{2}+\frac{\lambda }{2})_{i_{l-1}}(\frac{3}{4}+\frac{l}{2} +\frac{\lambda }{2})_{i_{l-1}}}{(-\alpha _l)_{i_{l-1}}(\alpha _l+l+\frac{1}{4}+\lambda )_{i_{l-1}}(1+\frac{l}{2}+\frac{\lambda }{2})_{i_l}(\frac{3}{4}+\frac{l}{2} +\frac{\lambda }{2})_{i_l}} \eta^{i_l}\nonumber\\
&=&  \int_{0}^{1} dt_l\;t_l^{\frac{l}{2}-\frac{5}{4}+\frac{\lambda }{2}} \int_{0}^{1} du_l\;u_l^{\frac{l}{2}-1+\frac{\lambda }{2}}\frac{1}{2\pi i} \oint dv_l\;\frac{1}{v_l} (1-\eta v_l (1-t_l)(1-u_l))^{-(l+\frac{1}{4}+\lambda )}\nonumber\\
&&\times \left(\frac{1- \frac{1}{v_l}}{1-\eta v_l (1-t_l)(1-u_l)}\right)^{\alpha_l} \left(\frac{t_lu_lv_l}{(v_l-1)}\frac{\eta }{1-\eta v_l (1-t_l)(1-u_l)}\right)^{i_{l-1}} \label{eq:68}
\end{eqnarray}
Substitute (\ref{eq:68}) into (\ref{eq:7}) where $l=1,2,3,\cdots$; apply $K_1$ into the second summation of sub-power series $y_1(z)$, apply $K_2$ into the third summation and $K_1$ into the second summation of sub-power series $y_2(z)$, apply $K_3$ into the forth summation, $K_2$ into the third summation and $K_1$ into the second summation of sub-power series $y_3(z)$, etc.\footnote{$y_1(z)$ means the sub-power series in (\ref{eq:7}) contains one term of $A_n's$, $y_2(z)$ means the sub-power series in (\ref{eq:7}) contains two terms of $A_n's$, $y_3(z)$ means the sub-power series in (\ref{eq:7}) contains three terms of $A_n's$, etc.} 
\begin{thm}
The general expression of the integral representation of the Lame polynomial in the algebraic form which makes $B_n$ term terminated is
\begin{eqnarray}
 y(z)&=& \sum_{n=0}^{\infty } y_{n}(z) = y_0(z)+ y_1(z)+ y_2(z)+y_3(z)+\cdots\nonumber\\
&=& c_0 z^{\lambda } \left\{ \sum_{i_0=0}^{\alpha _0}\frac{(-\alpha _0)_{i_0}(\alpha _0+\frac{1}{4}+\lambda )_{i_0}}{(1+\frac{\lambda }{2})_{i_0}(\frac{3}{4}+ \frac{\lambda }{2})_{i_0}}  \eta ^{i_0} \right.\nonumber\\
&&+ \sum_{n=1}^{\infty } \left\{\prod _{k=0}^{n-1} \left\{ \int_{0}^{1} dt_{n-k}\;t_{n-k}^{\frac{1}{2}(n-k-\frac{5}{2}+\lambda )} \int_{0}^{1} du_{n-k}\;u_{n-k}^{\frac{1}{2}(n-k-2+\lambda )} \right.\right. \nonumber\\
&&\times  \frac{1}{2\pi i}  \oint dv_{n-k} \frac{1}{v_{n-k}} \left( 1-\overleftrightarrow {w}_{n-k+1,n} v_{n-k}(1-t_{n-k})(1-u_{n-k})\right)^{-(n-k+\frac{1}{4}+\lambda )}\nonumber\\
&&\times \left(\frac{(v_{n-k}-1)}{v_{n-k}} \frac{1}{1-\overleftrightarrow {w}_{n-k+1,n}v_{n-k}(1-t_{n-k})(1-u_{n-k})}\right)^{\alpha _{n-k}}\nonumber\\
&&\times \left. \left( \overleftrightarrow {w}_{n-k,n}^{-\frac{1}{2}(n-k-1+\lambda )}\left(  \overleftrightarrow {w}_{n-k,n} \partial _{ \overleftrightarrow {w}_{n-k,n}}\right)^2 \overleftrightarrow {w}_{n-k,n}^{\frac{1}{2}(n-k-1+\lambda )} -\Omega _{n-k-1}^{(P)}\right) \right\}\nonumber\\
&&\times \left.\left. \sum_{i_0=0}^{\alpha _0}\frac{(-\alpha _0)_{i_0}(\alpha _0+\frac{1}{4}+\lambda )_{i_0}}{(1+\frac{\lambda }{2})_{i_0}(\frac{3}{4}+ \frac{\lambda }{2})_{i_0}}  \overleftrightarrow {w}_{1,n}^{i_0}\right\} \mu ^n \right\} \label{eq:69}
\end{eqnarray}
where
\begin{equation}\overleftrightarrow {w}_{i,j}=
\begin{cases} \displaystyle {\frac{1}{(v_i-1)}\; \frac{\overleftrightarrow w_{i+1,j}v_i t_i u_i}{1- \overleftrightarrow w_{i+1,j} v_i (1-t_i)(1-u_i)}}\;\;\mbox{where}\; i\leq j\cr
\eta \;\;\mbox{only}\;\mbox{if}\; i>j
\end{cases}\label{eq:70}
\end{equation}
and
\begin{equation}
\Omega _{n-k-1}^{(P)} =  \frac{a}{(2a-b-c)}\left( \left(\alpha _{n-k-1}+\frac{ n-k-1+\lambda }{2} \right) \left(\alpha _{n-k-1}+\frac{ n-k-\frac{1}{2}+\lambda }{2} \right) -\frac{q}{2^4 a} \right) \nonumber 
\end{equation}
In the above, the first sub-integral form contains one term of $A_n's$, the second one contains two terms of $A_n$'s, the third one contains three terms of $A_n$'s, etc.
\end{thm}
\begin{pot} 
According to (\ref{eq:7}), 
\begin{equation}
 y(z)= \sum_{n=0}^{\infty } y_{n}(z) = y_0(z)+ y_1(z)+ y_2(z)+y_3(z)+\cdots \label{eq:200}
\end{equation}
On the above, sub-power series $y_0(z) $, $y_1(z)$, $y_2(z)$ and $y_3(z)$ of Lame function in the algebraic form using 3TRF about $x=a$ are given by
\begin{subequations}
\begin{equation}
 y_0(z)= c_0 z^{\lambda } \sum_{i_0=0}^{\alpha _0} \frac{(-\alpha _0)_{i_0} (\alpha _0+ \frac{1}{4}+\lambda )_{i_0}}{(1+\frac{\lambda }{2})_{i_0}(\frac{3}{4} +\frac{\lambda }{2})_{i_0}} \eta ^{i_0}\label{eq:201a}
\end{equation}
\begin{eqnarray}
 y_1(z)&=& c_0 z^{\lambda } \Bigg\{ \sum_{i_0=0}^{\alpha _0} \frac{(i_0+\frac{\lambda }{2})^2- \frac{a}{(2a-b-c)}\left( (\alpha _0+\frac{\lambda }{2})(\alpha _0+\frac{1}{4}+\frac{\lambda }{2})-\frac{q}{2^4 a}\right)}{(i_0+\frac{1}{2}+\frac{\lambda }{2})(i_0+\frac{1}{4}+\frac{\lambda }{2})}\frac{(-\alpha _0)_{i_0} (\alpha _0+\frac{1}{4}+\lambda )_{i_0}}{(1+\frac{\lambda }{2})_{i_0}(\frac{3}{4}+ \frac{\lambda }{2})_{i_0}}\nonumber\\
&&\times \sum_{i_1=i_0}^{\alpha _1} \frac{(-\alpha _1)_{i_1} (\alpha _1+ \frac{5}{4}+\lambda )_{i_1}(\frac{3}{2}+\frac{\lambda}{2})_{i_0}(\frac{5}{4}+\frac{\lambda}{2})_{i_0}}{(-\alpha _1)_{i_0} (\alpha _1+ \frac{5}{4}+\lambda )_{i_0}(\frac{3}{2}+\frac{\lambda}{2})_{i_1}(\frac{5}{4}+\frac{\lambda}{2})_{i_1}} \eta ^{i_1} \Bigg\}\mu \label{eq:201b}
\end{eqnarray}
\begin{eqnarray}
 y_2(z) &=& c_0 z^{\lambda } \Bigg\{\sum_{i_0=0}^{\alpha _0} \frac{ (i_0+\frac{\lambda }{2})^2- \frac{a}{(2a-b-c)}\left( (\alpha _0+\frac{\lambda }{2})(\alpha _0+\frac{1}{4}+\frac{\lambda }{2})-\frac{q}{2^4 a}\right)}{(i_0+\frac{1}{2}+\frac{\lambda }{2})(i_0+\frac{1}{4}+\frac{\lambda }{2})}  \frac{(-\alpha _0)_{i_0} (\alpha _0+\frac{1}{4}+\lambda )_{i_0}}{(1+\frac{\lambda }{2})_{i_0}(\frac{3}{4}+ \frac{\lambda }{2})_{i_0}}  \nonumber\\
&&\times  \sum_{i_1=i_0}^{\alpha _1} \frac{ (i_1+\frac{1}{2}+\frac{\lambda }{2})^2- \frac{a}{(2a-b-c)}\left( (\alpha _1+\frac{1}{2}+\frac{\lambda }{2})(\alpha _1+\frac{3}{4}+\frac{\lambda }{2})-\frac{q}{2^4 a}\right)}{(i_1+1+\frac{\lambda }{2})(i_1+\frac{3}{4}+\frac{\lambda }{2})} \frac{(-\alpha _1)_{i_1} (\alpha _1+ \frac{5}{4}+\lambda )_{i_1}(\frac{3}{2}+\frac{\lambda}{2})_{i_0}(\frac{5}{4}+\frac{\lambda}{2})_{i_0}}{(-\alpha _1)_{i_0} (\alpha _1+ \frac{5}{4}+\lambda )_{i_0}(\frac{3}{2}+\frac{\lambda}{2})_{i_1}(\frac{5}{4}+\frac{\lambda}{2})_{i_1}}\nonumber\\
&&\times \sum_{i_2=i_1}^{\alpha _2} \frac{(-\alpha _2)_{i_2} (\alpha _2+ \frac{9}{4}+\lambda )_{i_2}(2+\frac{\lambda}{2})_{i_1}(\frac{7}{4}+\frac{\lambda}{2})_{i_1}}{(-\alpha _2)_{i_1} (\alpha _2+ \frac{9}{4}+\lambda )_{i_1}(2+\frac{\lambda}{2})_{i_2}(\frac{7}{4}+\frac{\lambda}{2})_{i_2}} \eta ^{i_2} \Bigg\} \mu ^2 \label{eq:201c}
\end{eqnarray}
\begin{eqnarray}
 y_3(z)&=& c_0 z^{\lambda } \Bigg\{\sum_{i_0=0}^{\alpha _0} \frac{ (i_0+\frac{\lambda }{2})^2- \frac{a}{(2a-b-c)}\left( (\alpha _0+\frac{\lambda }{2})(\alpha _0+\frac{1}{4}+\frac{\lambda }{2})-\frac{q}{2^4 a}\right)}{(i_0+\frac{1}{2}+\frac{\lambda }{2})(i_0+\frac{1}{4}+\frac{\lambda }{2})} \frac{(-\alpha _0)_{i_0} (\alpha _0+\frac{1}{4}+\lambda )_{i_0}}{(1+\frac{\lambda }{2})_{i_0}(\frac{3}{4}+ \frac{\lambda }{2})_{i_0}}  \nonumber\\
&&\times  \sum_{i_1=i_0}^{\alpha _1} \frac{ (i_1+\frac{1}{2}+\frac{\lambda }{2})^2- \frac{a}{(2a-b-c)}\left( (\alpha _1+\frac{1}{2}+\frac{\lambda }{2})(\alpha _1+\frac{3}{4}+\frac{\lambda }{2})-\frac{q}{2^4 a}\right)}{(i_1+1+\frac{\lambda }{2})(i_1+\frac{3}{4}+\frac{\lambda }{2})}    \frac{(-\alpha _1)_{i_1} (\alpha _1+ \frac{5}{4}+\lambda )_{i_1}(\frac{3}{2}+\frac{\lambda}{2})_{i_0}(\frac{5}{4}+\frac{\lambda}{2})_{i_0}}{(-\alpha _1)_{i_0} (\alpha _1+ \frac{5}{4}+\lambda )_{i_0}(\frac{3}{2}+\frac{\lambda}{2})_{i_1}(\frac{5}{4}+\frac{\lambda}{2})_{i_1}}\nonumber\\
&&\times \sum_{i_2=i_1}^{\alpha _2} \frac{ (i_2+1+\frac{\lambda }{2})^2- \frac{a}{(2a-b-c)}\left( (\alpha _2+1+\frac{\lambda }{2})(\alpha _2+\frac{5}{4}+\frac{\lambda }{2})-\frac{q}{2^4 a}\right)}{(i_2+\frac{3}{2}+\frac{\lambda }{2})(i_2+\frac{5}{4}+\frac{\lambda }{2})}  \frac{(-\alpha _2)_{i_2} (\alpha _2+ \frac{9}{4}+\lambda )_{i_2}(2+\frac{\lambda}{2})_{i_1}(\frac{7}{4}+\frac{\lambda}{2})_{i_1}}{(-\alpha _2)_{i_1} (\alpha _2+ \frac{9}{4}+\lambda )_{i_1}(2+\frac{\lambda}{2})_{i_2}(\frac{7}{4}+\frac{\lambda}{2})_{i_2}} \nonumber\\
&&\times \sum_{i_3=i_2}^{\alpha _3}  \frac{(-\alpha _3)_{i_3} (\alpha _3+ \frac{13}{4}+\lambda )_{i_3}(\frac{5}{2}+\frac{\lambda}{2})_{i_2}(\frac{9}{4}+\frac{\lambda}{2})_{i_2}}{(-\alpha _3)_{i_2} (\alpha _3+ \frac{13}{4}+\lambda )_{i_2}(\frac{5}{2}+\frac{\lambda}{2})_{i_3}(\frac{9}{4}+\frac{\lambda}{2})_{i_3}}\eta ^{i_3} \Bigg\} \mu ^3 \label{eq:201d}
\end{eqnarray}
\end{subequations}
Put $l=1$ in (\ref{eq:68}). Take the new (\ref{eq:68}) into (\ref{eq:201b}).
\begin{eqnarray}
 y_1(z)&=& c_0 z^{\lambda }\int_{0}^{1} dt_1\;t_1^{-\frac{3}{4}+\frac{\lambda }{2}} \int_{0}^{1} du_1\;u_1^{-\frac{1}{2}+\frac{\lambda }{2}} \nonumber\\
&&\times \frac{1}{2\pi i} \oint dv_1\;\frac{1}{v_1} \left(1- \eta v_1 (1-t_1)(1-u_1)\right)^{-(\frac{5}{4}+\lambda )} \left(\frac{(v_1-1)}{v_1}\frac{1}{1-\eta v_1 (1-t_1)(1-u_1)}\right)^{\alpha _1} \nonumber\\
&&\times  \left\{ \sum_{i_0=0}^{\alpha _0}\Bigg( \Big(i_0+\frac{\lambda }{2}\Big)^2 -\frac{a}{(2a-b-c)}\left( \Big(\alpha _0+\frac{\lambda }{2}\Big)\Big(\alpha _0+\frac{1}{4}+\frac{\lambda }{2}\Big)-\frac{q}{2^4 a}\right) \Bigg) \right.\nonumber\\
&&\times \left. \frac{(-\alpha _0)_{i_0} (\alpha _0+\frac{1}{4}+\lambda )_{i_0}}{(1+\frac{\lambda }{2})_{i_0}(\frac{3}{4}+ \frac{\lambda }{2})_{i_0}} \left(\frac{t_1 u_1 v_1}{(v_1-1)} \frac{\eta }{1-\eta v_1 (1-t_1)(1-u_1)}\right)^{i_0} \right\} \mu \nonumber\\
&=& c_0 z^{\lambda } \int_{0}^{1} dt_1\;t_1^{-\frac{3}{4}+\frac{\lambda }{2}} \int_{0}^{1} du_1\;u_1^{-\frac{1}{2}+\frac{\lambda }{2}} \nonumber\\
&&\times \frac{1}{2\pi i} \oint dv_1\;\frac{1}{v_1} \left(1- \eta v_1 (1-t_1)(1-u_1)\right)^{-(\frac{5}{4}+\lambda )} \left(\frac{(v_1-1)}{v_1}\frac{1}{1-\eta v_1 (1-t_1)(1-u_1)}\right)^{\alpha _1} \nonumber\\
&&\times  \Bigg( \overleftrightarrow {w}_{1,1}^{-\frac{\lambda }{2}}\left(  \overleftrightarrow {w}_{1,1} \partial _{ \overleftrightarrow {w}_{1,1}}\right)^2 \overleftrightarrow {w}_{1,1}^{\frac{\lambda }{2}}-\frac{a}{(2a-b-c)}\left( \Big(\alpha _0+\frac{\lambda }{2}\Big)\Big(\alpha _0+\frac{1}{4}+\frac{\lambda }{2}\Big)-\frac{q}{2^4 a}\right) \Bigg)\nonumber\\
&&\times \left\{ \sum_{i_0=0}^{\alpha _0} \frac{(-\alpha _0)_{i_0} (\alpha _0+\frac{1}{4}+\lambda )_{i_0}}{(1+\frac{\lambda }{2})_{i_0}(\frac{3}{4}+\frac{\lambda }{2})_{i_0}} \overleftrightarrow {w}_{1,1} ^{i_0} \right\} \mu \label{eq:202}
\end{eqnarray}
where 
\begin{equation}
 \overleftrightarrow {w}_{1,1} = \frac{t_1 u_1 v_1}{(v_1-1)}\; \frac{\eta }{1-\eta v_1 (1-t_1)(1-u_1)} \nonumber
\end{equation}
Put $l=2$ in (\ref{eq:68}). Take the new (\ref{eq:68}) into (\ref{eq:201c}). 
\begin{eqnarray}
y_2(z) &=& c_0 z^{\lambda } \int_{0}^{1} dt_2\;t_2^{-\frac{1}{4}+\frac{\lambda }{2}} \int_{0}^{1} du_2\;u_2^{\frac{\lambda }{2}} \nonumber\\
&&\times \frac{1}{2\pi i} \oint dv_2\;\frac{1}{v_2} \left(1- \eta v_2 (1-t_2)(1-u_2)\right)^{-(\frac{9}{4}+\lambda )}\left(\frac{(v_2-1)}{v_2}\frac{1}{1-\eta v_2(1-t_2)(1-u_2)}\right)^{\alpha_2} \nonumber\\
&&\times  \Bigg( \overleftrightarrow {w}_{2,2}^{-\frac{1}{2}(1+\lambda )}\left(  \overleftrightarrow {w}_{2,2} \partial _{ \overleftrightarrow {w}_{2,2}}\right)^2 \overleftrightarrow {w}_{2,2}^{\frac{1}{2}(1+\lambda )} -\frac{a}{(2a-b-c)}\left( \left(\alpha _1+\frac{1}{2}+\frac{\lambda }{2} \right)\left(\alpha _1+\frac{3}{4}+\frac{\lambda }{2}\right) -\frac{q}{2^4 a}\right) \Bigg)\nonumber\\
&&\times \left\{ \sum_{i_0=0}^{\alpha _0}\frac{ (i_0+\frac{\lambda }{2})^2-\frac{a}{(2a-b-c)}\left((\alpha _0+\frac{\lambda }{2})(\alpha _0+\frac{1}{4}+\frac{\lambda }{2})-\frac{q}{2^4 a}\right)}{(i_0+ \frac{1}{2}+ \frac{\lambda }{2})(i_0 + \frac{1}{4}+ \frac{\lambda }{2})} \frac{(-\alpha _0)_{i_0} (\alpha_0+\frac{1}{4}+\lambda )_{i_0}}{(1+\frac{\lambda }{2})_{i_0}(\frac{3}{4}+\frac{\lambda }{2})_{i_0}}\right.\nonumber\\
&&\times \left. \sum_{i_1=i_0}^{\alpha _1} \frac{(-\alpha _1)_{i_1}(\alpha _1+\frac{5}{4}+ \lambda)_{i_1}(\frac{3}{2}+\frac{\lambda }{2})_{i_0}(\frac{5}{4}+\frac{\lambda }{2})_{i_0}}{(-\alpha _1)_{i_0}(\alpha _1+\frac{5}{4}+ \lambda )_{i_0}(\frac{3}{2}+\frac{\lambda }{2})_{i_1}(\frac{5}{4} +\frac{\lambda }{2})_{i_1}} \overleftrightarrow {w}_{2,2}^{i_1} \right\} \mu^2 \label{eq:203}
\end{eqnarray}
where
\begin{equation}
 \overleftrightarrow {w}_{2,2} = \frac{t_2 u_2 v_2}{(v_2-1)} \;\frac{\eta }{1-\eta v_2 (1-t_2)(1-u_2)}\nonumber
\end{equation} 
Put $l=1$ and $\eta =\overleftrightarrow {w}_{2,2}$ in (\ref{eq:68}). Take the new (\ref{eq:68}) into (\ref{eq:203}).
\begin{eqnarray}
y_2(z) &=& c_0 z^{\lambda } \int_{0}^{1} dt_2\;t_2^{-\frac{1}{4}+\frac{\lambda }{2}} \int_{0}^{1} du_2\;u_2^{\frac{\lambda }{2}} \nonumber\\
&&\times \frac{1}{2\pi i} \oint dv_2\;\frac{1}{v_2} \left(1- \eta v_2 (1-t_2)(1-u_2)\right)^{-(\frac{9}{4}+\lambda )}\left(\frac{(v_2-1)}{v_2}\frac{1}{1-\eta v_2(1-t_2)(1-u_2)}\right)^{\alpha_2} \nonumber\\
&&\times  \Bigg( \overleftrightarrow {w}_{2,2}^{-\frac{1}{2}(1+\lambda )}\left(  \overleftrightarrow {w}_{2,2} \partial _{ \overleftrightarrow {w}_{2,2}}\right)^2 \overleftrightarrow {w}_{2,2}^{\frac{1}{2}(1+\lambda )} -\frac{a}{(2a-b-c)}\left(\left(\alpha _1+\frac{1}{2}+\frac{\lambda }{2} \right)\left(\alpha _1+\frac{3}{4}+\frac{\lambda }{2}\right) -\frac{q}{2^4 a}\right)\Bigg)\nonumber\\
&&\times \int_{0}^{1} dt_1\;t_1^{-\frac{3}{4}+\frac{\lambda }{2}} \int_{0}^{1} du_1\;u_1^{-\frac{1}{2}+\frac{\lambda }{2}}\nonumber\\
&&\times \frac{1}{2\pi i} \oint dv_1\;\frac{1}{v_1}(1-\overleftrightarrow {w}_{2,2} v_1 (1-t_1)(1-u_1))^{-(\frac{5}{4}+\lambda )} \left(\frac{(v_1-1)}{v_1}\frac{1}{1-\overleftrightarrow {w}_{2,2} v_1 (1-t_1)(1-u_1)}\right)^{\alpha_1} \nonumber\\
&&\times \Bigg( \overleftrightarrow {w}_{1,2}^{-\frac{\lambda }{2}}\left(  \overleftrightarrow {w}_{1,2} \partial _{ \overleftrightarrow {w}_{1,2}}\right)^2 \overleftrightarrow {w}_{1,2}^{\frac{\lambda }{2}} -\frac{a}{(2a-b-c)}\left( \left(\alpha _0+\frac{\lambda }{2} \right)\left(\alpha _0+\frac{1}{4}+\frac{\lambda }{2}\right)-\frac{q}{2^4 a}\right)\Bigg) \nonumber\\
&&\times \left\{ \sum_{i_0=0}^{\alpha _0} \frac{(-\alpha _0)_{i_0} (\alpha _0+\frac{1}{4}+\lambda )_{i_0}}{(1+\frac{\lambda }{2})_{i_0}(\frac{3}{4}+\frac{\lambda }{2})_{i_0}} \overleftrightarrow {w}_{1,2} ^{i_0} \right\} \mu^2 \label{eq:204}
\end{eqnarray}
where
\begin{equation}
 \overleftrightarrow {w}_{1,2}=\frac{t_1 u_1 v_1}{(v_1-1)}\; \frac{\overleftrightarrow {w}_{2,2}}{1-\overleftrightarrow {w}_{2,2} v_1 (1-t_1)(1-u_1)}\nonumber
\end{equation} 
By using similar process for the previous cases of integral forms of $y_1(z)$ and $y_2(z)$, the integral form of sub-power series expansion of $y_3(z)$ is
\begin{eqnarray}
y_3(z)&=& c_0 z^{\lambda } \int_{0}^{1} dt_3\;t_3^{\frac{1}{4}+\frac{\lambda }{2}} \int_{0}^{1} du_3\;u_3^{\frac{1}{2}+\frac{\lambda }{2}}\nonumber\\
&&\times \frac{1}{2\pi i} \oint dv_3\;\frac{1}{v_3} (1-\eta v_3 (1-t_3)(1-u_3))^{-(\frac{13}{4}+\lambda )} \left(\frac{(v_3-1)}{v_3}\frac{1}{1-\eta v_3 (1-t_3)(1-u_3)}\right)^{\alpha_3}\nonumber\\
&&\times \Bigg( \overleftrightarrow {w}_{3,3}^{-\frac{1}{2}(2+\lambda )}\left(  \overleftrightarrow {w}_{3,3} \partial _{ \overleftrightarrow {w}_{3,3}}\right)^2 \overleftrightarrow {w}_{3,3}^{\frac{1}{2}(2+\lambda )} -\frac{a}{(2a-b-c)}\left(\left( \alpha _2+1+\frac{\lambda }{2}\right) \left( \alpha _2+\frac{5}{4}+\frac{\lambda }{2}\right) -\frac{q}{2^4 a}\right)\Bigg)\nonumber\\
&&\times  \int_{0}^{1} dt_2\;t_2^{-\frac{1}{4}+\frac{\lambda }{2}} \int_{0}^{1} du_2\;u_2^{\frac{\lambda }{2}}  \nonumber\\
&&\times\frac{1}{2\pi i} \oint dv_2\;\frac{1}{v_2} (1-\overleftrightarrow {w}_{3,3} v_2 (1-t_2)(1-u_2))^{-(\frac{9}{4}+\lambda )} \left(\frac{(v_2-1)}{v_2}\frac{1}{1-\overleftrightarrow {w}_{3,3} v_2 (1-t_2)(1-u_2)}\right)^{\alpha_2} \nonumber\\
&&\times \Bigg( \overleftrightarrow {w}_{2,3}^{-\frac{1}{2}(1+\lambda )}\left(  \overleftrightarrow {w}_{2,3} \partial _{ \overleftrightarrow {w}_{2,3}}\right)^2 \overleftrightarrow {w}_{2,3}^{\frac{1}{2}(1+\lambda )} -\frac{a}{(2a-b-c)}\left( \left( \alpha _1+\frac{1}{2}+\frac{\lambda }{2}\right) \left( \alpha _1+\frac{3}{4}+\frac{\lambda }{2}\right)-\frac{q}{2^4 a}\right)\Bigg)\nonumber\\
&&\times \int_{0}^{1} dt_1\;t_1^{-\frac{3}{4}+\frac{\lambda }{2}} \int_{0}^{1} du_1\;u_1^{-\frac{1}{2}+\frac{\lambda }{2}} \nonumber\\
&&\times \frac{1}{2\pi i} \oint dv_1\;\frac{1}{v_1} (1-\overleftrightarrow w_{2,3} v_1 (1-t_1)(1-u_1))^{-(\frac{5}{4}+\lambda )}\left(\frac{(v_1-1)}{v_1}\frac{1}{1-\overleftrightarrow w_{2,3} v_1 (1-t_1)(1-u_1)}\right)^{\alpha_1} \nonumber\\
&&\times \Bigg( \overleftrightarrow {w}_{1,3}^{-\frac{\lambda }{2}}\left(  \overleftrightarrow {w}_{1,3} \partial _{ \overleftrightarrow {w}_{1,3}}\right)^2 \overleftrightarrow {w}_{1,3}^{\frac{\lambda }{2}} -\frac{a}{(2a-b-c)}\left( \left( \alpha _0+\frac{\lambda }{2}\right) \left( \alpha _0+\frac{1}{4}+\frac{\lambda }{2}\right)-\frac{q}{2^4 a}\right)\Bigg) \nonumber\\
&&\times  \left\{ \sum_{i_0=0}^{\alpha _0} \frac{(-\alpha _0)_{i_0} (\alpha _0+\frac{1}{4}+\lambda )_{i_0}}{(1+\frac{\lambda }{2})_{i_0}(\frac{3}{4}+ \frac{\lambda }{2})_{i_0}} \overleftrightarrow{w}_{1,3}^{i_0} \right\} \mu^3 \label{eq:205}
\end{eqnarray}
where
\begin{equation}
\begin{cases} \overleftrightarrow {w}_{3,3} = \frac{t_3 u_3 v_3}{(v_3-1)}\; \frac{\eta }{1- \eta v_3 (1-t_3)(1-u_3)}  \cr
\overleftrightarrow {w}_{2,3} = \frac{t_2 u_2 v_2}{(v_2-1)}\; \frac{\overleftrightarrow{w}_{3,3} }{1- \overleftrightarrow{w}_{3,3} v_2 (1-t_2)(1-u_2)} \cr
\overleftrightarrow w_{1,3}= \frac{t_1 u_1 v_1}{(v_1-1)}\frac{\overleftrightarrow w_{2,3} }{1-\overleftrightarrow w_{2,3} v_1 (1-t_1)(1-u_1)}
\end{cases}
\nonumber
\end{equation}
By repeating this process for all higher terms of integral forms of sub-summation $y_m(z)$ terms where $m \geq 4$, we obtain every integral forms of $y_m(z)$ terms. 
Since we substitute (\ref{eq:201a}), (\ref{eq:202}), (\ref{eq:204}), (\ref{eq:205}) and including all integral forms of $y_m(x)$ terms where $m \geq 4$ into (\ref{eq:200}), we obtain (\ref{eq:69}).
\qed
\end{pot} 
Take $c_0$= 1 as $\lambda =0$  for the first independent solution of Lame equation and $\lambda =\frac{1}{2}$ for the second one into (\ref{eq:69}). 
\begin{rmk}
The integral representation of the first kind of Lame polynomial which makes $B_n$ term terminated about $x=a$  as $\alpha = 2(2\alpha_j +j) $ or $ -2(2\alpha_j +j)-1$ where $j, \alpha _j =0,1,2,\cdots$ is
\begin{eqnarray}
 y(z)&=& LF_{\alpha _j}\left( a, b, c, q, \alpha = 2(2\alpha_j +j)\; \mbox{or} -2(2\alpha_j +j)-1; z= x-a, \mu = \frac{-(2a-b-c)z}{(a-b)(a-c)}, \eta = \frac{-z^2}{(a-b)(a-c)} \right) \nonumber\\
&=& _2F_1 \left(-\alpha _0, \alpha _0+\frac{1}{4};\frac{3}{4}; \eta \right) + \sum_{n=1}^{\infty } \Bigg\{\prod _{k=0}^{n-1} \Bigg\{ \int_{0}^{1} dt_{n-k}\;t_{n-k}^{\frac{1}{2}(n-k-\frac{5}{2})} \int_{0}^{1} du_{n-k}\;u_{n-k}^{\frac{1}{2}(n-k-2)} \nonumber\\
&&\times  \frac{1}{2\pi i}  \oint dv_{n-k} \frac{1}{v_{n-k}} \left( 1-\overleftrightarrow {w}_{n-k+1,n} v_{n-k}(1-t_{n-k})(1-u_{n-k})\right)^{-(n-k+\frac{1}{4})}\nonumber\\
&&\times \left(\frac{(v_{n-k}-1)}{v_{n-k}} \frac{1}{1-\overleftrightarrow {w}_{n-k+1,n}v_{n-k}(1-t_{n-k})(1-u_{n-k})}\right)^{\alpha _{n-k}}\nonumber\\
&&\times \left( \overleftrightarrow {w}_{n-k,n}^{-\frac{1}{2}(n-k-1)}\left(  \overleftrightarrow {w}_{n-k,n} \partial _{ \overleftrightarrow {w}_{n-k,n}}\right)^2 \overleftrightarrow {w}_{n-k,n}^{\frac{1}{2}(n-k-1)} -\Omega _{n-k-1}^{(P)}\right) \Bigg\} \; _2F_1 \left(-\alpha _0, \alpha _0+\frac{1}{4};\frac{3}{4};  \overleftrightarrow {w}_{1,n}\right) \Bigg\} \mu ^n \nonumber
\end{eqnarray}
where
\begin{equation}
\Omega _{n-k-1}^{(P)} =  \frac{a}{(2a-b-c)}\left( \left(\alpha _{n-k-1}+\frac{ n-k-1 }{2} \right) \left(\alpha _{n-k-1}+\frac{ n-k-\frac{1}{2} }{2} \right) -\frac{q}{2^4 a} \right) \nonumber 
\end{equation}
\end{rmk}
\begin{rmk}
The integral representation of the second kind of Lame polynomial which makes $B_n$ term terminated about $x=a$  as $\alpha = 2(2\alpha_j +j )+1$ or $-2(2\alpha _j+j+1)$ where $j, \alpha _j =0,1,2,\cdots$ is
\begin{eqnarray}
y(z)&=& LS_{\alpha _j}\Bigg(  a, b, c, q, \alpha = 2(2\alpha_j +j)+1 \;\mbox{or} -2(2\alpha_j +j+1); z= x-a, \mu = \frac{-(2a-b-c)z}{(a-b)(a-c)}, \eta = \frac{-z^2}{(a-b)(a-c)} \Bigg) \nonumber\\
&=& z^{\frac{1}{2}} \Bigg\{\; _2F_1 \left(-\alpha _0, \alpha _0+\frac{3}{4};\frac{5}{4}; \eta \right)
+  \sum_{n=1}^{\infty } \Bigg\{\prod _{k=0}^{n-1} \Bigg\{ \int_{0}^{1} dt_{n-k}\;t_{n-k}^{\frac{1}{2}(n-k-2)} \int_{0}^{1} du_{n-k}\;u_{n-k}^{\frac{1}{2}(n-k-\frac{3}{2})}  \nonumber\\
&&\times  \frac{1}{2\pi i}  \oint dv_{n-k} \frac{1}{v_{n-k}} \left( 1-\overleftrightarrow {w}_{n-k+1,n} v_{n-k}(1-t_{n-k})(1-u_{n-k})\right)^{-(n-k+\frac{3}{4})}\nonumber\\
&&\times \left(\frac{(v_{n-k}-1)}{v_{n-k}} \frac{1}{1-\overleftrightarrow {w}_{n-k+1,n}v_{n-k}(1-t_{n-k})(1-u_{n-k})}\right)^{\alpha _{n-k}}\nonumber\\
&&\times   \Bigg( \overleftrightarrow {w}_{n-k,n}^{-\frac{1}{2}(n-k-\frac{1}{2})}\left(  \overleftrightarrow {w}_{n-k,n} \partial _{ \overleftrightarrow {w}_{n-k,n}}\right)^2 \overleftrightarrow {w}_{n-k,n}^{\frac{1}{2}(n-k-\frac{1}{2})} -\Omega _{n-k-1}^{(P)} \Bigg) \Bigg\}\; _2F_1 \left(-\alpha _0,  \alpha _0+\frac{3}{4};\frac{5}{4}; \overleftrightarrow {w}_{1,n}\right)  \Bigg\} \mu ^n \Bigg\}\nonumber 
\end{eqnarray}
and
\begin{equation}
\Omega _{n-k-1}^{(P)} =  \frac{a}{(2a-b-c)}\left( \left(\alpha _{n-k-1}+\frac{ n-k-\frac{1}{2}}{2} \right) \left(\alpha _{n-k-1}+\frac{ n-k }{2} \right) -\frac{q}{2^4 a} \right) \nonumber 
\end{equation}
\end{rmk}
\subsection{Infinite series}
Let's consider the integral representation of Lame equation about $x=a$ for infinite series by applying 3TRF.
There is a generalized hypergeometric function which is written by
\begin{eqnarray}
M_l &=& \sum_{i_l= i_{l-1}}^{\infty } \frac{\left(-\frac{\alpha }{4}+\frac{l}{2}+\frac{\lambda }{2}\right)_{i_l}\left( \frac{\alpha }{4}+\frac{1}{4}+\frac{l}{2}+\frac{\lambda }{2}\right)_{i_l}(1+\frac{l}{2}+\frac{\lambda }{2})_{i_{l-1}}(\frac{3}{4}+\frac{l}{2} +\frac{\lambda }{2})_{i_{l-1}}}{\left(-\frac{\alpha }{4}+\frac{l}{2}+\frac{\lambda }{2}\right)_{i_{l-1}}\left( \frac{\alpha }{4}+\frac{1}{4}+\frac{l}{2}+\frac{\lambda }{2}\right)_{i_{l-1}}(1+\frac{l}{2}+\frac{\lambda }{2})_{i_l}(\frac{3}{4}+\frac{l}{2} +\frac{\lambda }{2})_{i_l}} \eta^{i_l} \label{er:63}\\
&=& \eta^{i_{l-1}} 
\sum_{j=0}^{\infty } \frac{B(i_{l-1}+\frac{l}{2}-\frac{1}{4}+\frac{\lambda }{2},j+1) B(i_{l-1}+\frac{l}{2}+\frac{\lambda }{2},j+1)\left(-\frac{\alpha }{4}+\frac{l}{2}+\frac{\lambda }{2}+i_{l-1}\right)_j \left( \frac{\alpha }{4}+\frac{1}{4}+\frac{l}{2}+\frac{\lambda }{2}+i_{l-1} \right)_j}{(i_{l-1}+\frac{l}{2}-\frac{1}{4}+\frac{\lambda }{2})^{-1}(i_{l-1}+\frac{l}{2}+ \frac{\lambda }{2})^{-1}(1)_j \;j!} \eta^j \nonumber
\end{eqnarray}
Substitute (\ref{eq:64a}) and (\ref{eq:64b}) into (\ref{er:63}). And divide $(i_{l-1}+\frac{l}{2}-\frac{1}{4}+\frac{\lambda }{2})(i_{l-1}+\frac{l}{2}+ \frac{\lambda }{2})$ into the new (\ref{er:63}).
\begin{eqnarray}
V_l &=& \frac{1}{(i_{l-1}+\frac{l}{2}-\frac{1}{4}+\frac{\lambda }{2})(i_{l-1}+\frac{l}{2}+ \frac{\lambda }{2})}\nonumber\\
&&\times \sum_{i_l= i_{l-1}}^{\infty } \frac{\left(-\frac{\alpha }{4}+\frac{l}{2}+\frac{\lambda }{2}\right)_{i_l}\left( \frac{\alpha }{4}+\frac{1}{4}+\frac{l}{2}+\frac{\lambda }{2}\right)_{i_l}(1+\frac{l}{2}+\frac{\lambda }{2})_{i_{l-1}}(\frac{3}{4}+\frac{l}{2} +\frac{\lambda }{2})_{i_{l-1}}}{\left(-\frac{\alpha }{4}+\frac{l}{2}+\frac{\lambda }{2}\right)_{i_{l-1}}\left( \frac{\alpha }{4}+\frac{1}{4}+\frac{l}{2}+\frac{\lambda }{2}\right)_{i_{l-1}}(1+\frac{l}{2}+\frac{\lambda }{2})_{i_l}(\frac{3}{4}+\frac{l}{2} +\frac{\lambda }{2})_{i_l}} \eta^{i_l} \nonumber\\
&=&  \int_{0}^{1} dt_l\;t_l^{\frac{l}{2}-\frac{5}{4}+\frac{\lambda }{2}} \int_{0}^{1} du_l\;u_l^{\frac{l}{2}-1+\frac{\lambda }{2}} (t_l u_l\eta )^{i_{l-1}}\nonumber\\
&&\times \sum_{j=0}^{\infty } \frac{\left(-\frac{\alpha }{4}+\frac{l}{2}+\frac{\lambda }{2}+i_{l-1}\right)_j \left( \frac{\alpha }{4}+\frac{1}{4}+\frac{l}{2}+\frac{\lambda }{2}+i_{l-1} \right)_j}{(1)_j \;j!} (\eta (1-t_l)(1-u_l))^j
\label{er:65}
\end{eqnarray}
The hypergeometric function is defined by
\begin{eqnarray}
_2F_1 \left( \alpha ,\beta ; \gamma ; z \right) &=& \sum_{n=0}^{\infty } \frac{(\alpha )_n (\beta )_n}{(\gamma )_n (n!)} z^n \nonumber\\
&=&  \frac{1}{2\pi i} \frac{\Gamma( 1+\alpha  -\gamma )}{\Gamma (\alpha )} \int_0^{(1+)} dv_l\; (-1)^{\gamma }(-v_l)^{\alpha -1} (1-v_l )^{\gamma -\alpha -1} (1-zv_l)^{-\beta }\hspace{1cm}\label{er:66}\\
&& \mbox{where} \;\gamma -\alpha  \ne 1,2,3,\cdots, \;\mbox{Re}(\alpha )>0 \nonumber
\end{eqnarray}
Replace $\alpha $, $\beta $, $\gamma $ and $z$ by $-\frac{\alpha }{4}+\frac{l}{2}+\frac{\lambda }{2}+i_{l-1}$, $ \frac{\alpha }{4}+\frac{1}{4}+\frac{l}{2}+\frac{\lambda }{2}+i_{l-1}$, 1 and $\eta (1-t_l)(1-u_l)$ in (\ref{er:66}). Take the new (\ref{er:66}) into (\ref{er:65})
\begin{eqnarray}
V_l &=& \frac{1}{(i_{l-1}+\frac{l}{2}-\frac{1}{4}+\frac{\lambda }{2})(i_{l-1}+\frac{l}{2}+ \frac{\lambda }{2})}\nonumber\\
&&\times \sum_{i_l= i_{l-1}}^{\infty } \frac{\left(-\frac{\alpha }{4}+\frac{l}{2}+\frac{\lambda }{2}\right)_{i_l}\left( \frac{\alpha }{4}+\frac{1}{4}+\frac{l}{2}+\frac{\lambda }{2}\right)_{i_l}(1+\frac{l}{2}+\frac{\lambda }{2})_{i_{l-1}}(\frac{3}{4}+\frac{l}{2} +\frac{\lambda }{2})_{i_{l-1}}}{\left(-\frac{\alpha }{4}+\frac{l}{2}+\frac{\lambda }{2}\right)_{i_{l-1}}\left( \frac{\alpha }{4}+\frac{1}{4}+\frac{l}{2}+\frac{\lambda }{2}\right)_{i_{l-1}}(1+\frac{l}{2}+\frac{\lambda }{2})_{i_l}(\frac{3}{4}+\frac{l}{2} +\frac{\lambda }{2})_{i_l}} \eta^{i_l} \nonumber\\
&=&  \int_{0}^{1} dt_l\;t_l^{\frac{l}{2}-\frac{5}{4}+\frac{\lambda }{2}} \int_{0}^{1} du_l\;u_l^{\frac{l}{2}-1+\frac{\lambda }{2}}\frac{1}{2\pi i} \oint dv_l\;\frac{1}{v_l} \left(\frac{v_l-1}{v_l}\right)^{\frac{1}{2}(\frac{\alpha }{2}-l-\lambda )} (1-\eta v_l (1-t_l)(1-u_l))^{-\frac{1}{2}(\frac{\alpha }{2}+\frac{1}{2}+l+\lambda )}\nonumber\\
&&\times  \left(\frac{t_lu_lv_l}{(v_l-1)}\frac{\eta }{1-\eta v_l (1-t_l)(1-u_l)}\right)^{i_{l-1}} \label{er:67}
\end{eqnarray}
Substitute (\ref{er:67}) into (\ref{eq:13}) where $l=1,2,3,\cdots$; apply $V_1$ into the second summation of sub-power series $y_1(z)$, apply $V_2$ into the third summation and $V_1$ into the second summation of sub-power series $y_2(z)$, apply $V_3$ into the forth summation, $V_2$ into the third summation and $V_1$ into the second summation of sub-power series $y_3(z)$, etc.\footnote{$y_1(z)$ means the sub-power series in (\ref{eq:13}) contains one term of $A_n's$, $y_2(z)$ means the sub-power series in (\ref{eq:13}) contains two terms of $A_n's$, $y_3(z)$ means the sub-power series in (\ref{eq:13}) contains three terms of $A_n's$, etc.} 
\begin{thm}
The general expression of the integral representation of the Lame equation about $x=a$ for infinite series is given by
\begin{eqnarray}
 y(z)&=& \sum_{n=0}^{\infty } y_{n}(z) = y_0(z)+ y_1(z)+ y_2(z)+y_3(z)+\cdots \nonumber\\
&=& c_0 z^{\lambda } \left\{ \sum_{i_0=0}^{\infty }\frac{(-\frac{\alpha }{4}+\frac{\lambda }{2})_{i_0}(\frac{\alpha }{4}+\frac{1}{4}+\frac{\lambda }{2})_{i_0}}{(1+\frac{\lambda }{2})_{i_0}(\frac{3}{4}+ \frac{\lambda }{2})_{i_0}}  \eta ^{i_0}\right. \nonumber\\
&&+ \sum_{n=1}^{\infty } \left\{\prod _{k=0}^{n-1} \Bigg\{ \int_{0}^{1} dt_{n-k}\;t_{n-k}^{\frac{1}{2}(n-k-\frac{5}{2}+\lambda )} \int_{0}^{1} du_{n-k}\;u_{n-k}^{\frac{1}{2}(n-k-2+\lambda )} \right.\nonumber\\
&&\times  \frac{1}{2\pi i}  \oint dv_{n-k} \frac{1}{v_{n-k}} \left(\frac{ v_{n-k}-1 }{v_{n-k}} \right)^{\frac{1}{2}(\frac{\alpha }{2}-n+k-\lambda )}   \left( 1-\overleftrightarrow {w}_{n-k+1,n} v_{n-k}(1-t_{n-k})(1-u_{n-k})\right)^{-\frac{1}{2}(\frac{\alpha }{2}+\frac{1}{2}+n-k+\lambda )} \nonumber\\
&&\times \Bigg( \overleftrightarrow {w}_{n-k,n}^{-\frac{1}{2}(n-k-1+\lambda )}\left(  \overleftrightarrow {w}_{n-k,n} \partial _{ \overleftrightarrow {w}_{n-k,n}}\right)^2 \overleftrightarrow {w}_{n-k,n}^{\frac{1}{2}(n-k-1+\lambda )}-\Gamma ^{(I)}\Bigg) \Bigg\}\nonumber\\
&&\times \left.\left. \sum_{i_0=0}^{\infty }\frac{(-\frac{\alpha }{4}+\frac{\lambda }{2})_{i_0}(\frac{\alpha }{4}+\frac{1}{4}+\frac{\lambda }{2})_{i_0}}{(1+\frac{\lambda }{2})_{i_0}(\frac{3}{4}+ \frac{\lambda }{2})_{i_0}} \overleftrightarrow {w}_{1,n}^{i_0}\right\} \mu ^n \right\} \label{eq:73}
\end{eqnarray}
where
\begin{equation}
\Gamma ^{(I)} = \frac{a}{2^4(2a-b-c)}\left( \alpha (\alpha +1) -\frac{q}{a}\right) \nonumber
\end{equation}
In the above, the first sub-integral form contains one term of $A_n's$, the second one contains two terms of $A_n$'s, the third one contains three terms of $A_n$'s, etc.
\end{thm}
\begin{pot} 
In (\ref{eq:13}) sub-power series $y_0(z) $, $y_1(z)$, $y_2(z)$ and $y_3(z)$ of Lame equation for infinite series about $x=a$ using 3TRF are given by
\begin{subequations}
\begin{equation}
 y_0(z)= c_0 z^{\lambda } \sum_{i_0=0}^{\infty } \frac{\left(-\frac{\alpha }{4} +\frac{\lambda }{2}\right)_{i_0} \left( \frac{\alpha }{4}+\frac{1}{4}+\frac{\lambda }{2}\right)_{i_0}}{(1+\frac{\lambda }{2})_{i_0}(\frac{3}{4} +\frac{\lambda }{2})_{i_0}} \eta ^{i_0}\label{er:68a}
\end{equation}
\begin{eqnarray}
 y_1(z)&=& c_0 z^{\lambda } \left\{ \sum_{i_0=0}^{\infty} \frac{ (i_0+\frac{\lambda }{2})^2- \Gamma ^{(I)}}{(i_0+\frac{1}{2}+\frac{\lambda }{2})(i_0+\frac{1}{4}+\frac{\lambda }{2})}\frac{\left(-\frac{\alpha }{4} +\frac{\lambda }{2}\right)_{i_0} \left( \frac{\alpha }{4}+\frac{1}{4}+\frac{\lambda }{2}\right)_{i_0}}{(1+\frac{\lambda }{2})_{i_0}(\frac{3}{4} +\frac{\lambda }{2})_{i_0}}\right. \nonumber\\
&&\times \left.\sum_{i_1=i_0}^{\infty} \frac{\left(-\frac{\alpha }{4} +\frac{1}{2}+\frac{\lambda }{2}\right)_{i_1} \left( \frac{\alpha }{4} +\frac{3}{4}+\frac{\lambda }{2}\right)_{i_1}(\frac{3}{2}+\frac{\lambda}{2})_{i_0}(\frac{5}{4}+\frac{\lambda}{2})_{i_0}}{\left(-\frac{\alpha }{4} +\frac{1}{2}+\frac{\lambda }{2}\right)_{i_0} \left( \frac{\alpha }{4} +\frac{3}{4}+\frac{\lambda }{2}\right)_{i_0}(\frac{3}{2}+\frac{\lambda}{2})_{i_1}(\frac{5}{4}+\frac{\lambda}{2})_{i_1}} \eta ^{i_1} \right\}\mu \label{er:68b}
\end{eqnarray}
\begin{eqnarray}
 y_2(z) &=& c_0 z^{\lambda } \left\{\sum_{i_0=0}^{\infty} \frac{ (i_0+\frac{\lambda }{2})^2- \Gamma ^{(I)}}{(i_0+\frac{1}{2}+\frac{\lambda }{2})(i_0+\frac{1}{4}+\frac{\lambda }{2})}\frac{\left(-\frac{\alpha }{4} +\frac{\lambda }{2}\right)_{i_0} \left( \frac{\alpha }{4}+\frac{1}{4}+\frac{\lambda }{2}\right)_{i_0}}{(1+\frac{\lambda }{2})_{i_0}(\frac{3}{4} +\frac{\lambda }{2})_{i_0}}  \right. \nonumber\\
&&\times  \sum_{i_1=i_0}^{\infty} \frac{ (i_1+\frac{1}{2}+\frac{\lambda }{2})^2- \Gamma ^{(I)}}{(i_1+1+\frac{\lambda }{2})(i_1+\frac{3}{4}+\frac{\lambda }{2})} \frac{\left(-\frac{\alpha }{4} +\frac{1}{2}+\frac{\lambda }{2}\right)_{i_1} \left( \frac{\alpha }{4} +\frac{3}{4}+\frac{\lambda }{2}\right)_{i_1}(\frac{3}{2}+\frac{\lambda}{2})_{i_0}(\frac{5}{4}+\frac{\lambda}{2})_{i_0}}{\left(-\frac{\alpha }{4} +\frac{1}{2}+\frac{\lambda }{2}\right)_{i_0} \left( \frac{\alpha }{4} +\frac{3}{4}+\frac{\lambda }{2}\right)_{i_0}(\frac{3}{2}+\frac{\lambda}{2})_{i_1}(\frac{5}{4}+\frac{\lambda}{2})_{i_1}}\nonumber\\
&&\times \left. \sum_{i_2=i_1}^{\infty} \frac{\left(-\frac{\alpha }{4} +1+\frac{\lambda }{2}\right)_{i_2} \left( \frac{\alpha }{4} +\frac{5}{4}+\frac{\lambda }{2}\right)_{i_2}(2+\frac{\lambda}{2})_{i_1}(\frac{7}{4}+\frac{\lambda}{2})_{i_1}}{\left(-\frac{\alpha }{4} +1+\frac{\lambda }{2}\right)_{i_1} \left( \frac{\alpha }{4} +\frac{5}{4}+\frac{\lambda }{2}\right)_{i_1}(2+\frac{\lambda}{2})_{i_2}(\frac{7}{4}+\frac{\lambda}{2})_{i_2}} \eta ^{i_2} \right\} \mu ^2 \label{er:68c}
\end{eqnarray}
\begin{eqnarray}
 y_3(z)&=& c_0 z^{\lambda } \left\{\sum_{i_0=0}^{\infty} \frac{ (i_0+\frac{\lambda }{2})^2- \Gamma ^{(I)}}{(i_0+\frac{1}{2}+\frac{\lambda }{2})(i_0+\frac{1}{4}+\frac{\lambda }{2})}\frac{\left(-\frac{\alpha }{4} +\frac{\lambda }{2}\right)_{i_0} \left( \frac{\alpha }{4}+\frac{1}{4}+\frac{\lambda }{2}\right)_{i_0}}{(1+\frac{\lambda }{2})_{i_0}(\frac{3}{4} +\frac{\lambda }{2})_{i_0}}  \right. \nonumber\\
&&\times  \sum_{i_1=i_0}^{\infty} \frac{ (i_1+\frac{1}{2}+\frac{\lambda }{2})^2- \Gamma ^{(I)}}{(i_1+1+\frac{\lambda }{2})(i_1+\frac{3}{4}+\frac{\lambda }{2})} \frac{\left(-\frac{\alpha }{4} +\frac{1}{2}+\frac{\lambda }{2}\right)_{i_1} \left( \frac{\alpha }{4} +\frac{3}{4}+\frac{\lambda }{2}\right)_{i_1}(\frac{3}{2}+\frac{\lambda}{2})_{i_0}(\frac{5}{4}+\frac{\lambda}{2})_{i_0}}{\left(-\frac{\alpha }{4} +\frac{1}{2}+\frac{\lambda }{2}\right)_{i_0} \left( \frac{\alpha }{4} +\frac{3}{4}+\frac{\lambda }{2}\right)_{i_0}(\frac{3}{2}+\frac{\lambda}{2})_{i_1}(\frac{5}{4}+\frac{\lambda}{2})_{i_1}}\nonumber\\
&&\times \sum_{i_2=i_1}^{\infty} \frac{ (i_2+1+\frac{\lambda }{2})^2- \Gamma ^{(I)}}{(i_2+\frac{3}{2}+\frac{\lambda }{2})(i_2+\frac{5}{4}+\frac{\lambda }{2})}  \frac{\left(-\frac{\alpha }{4} +1+\frac{\lambda }{2}\right)_{i_2} \left( \frac{\alpha }{4} +\frac{5}{4}+\frac{\lambda }{2}\right)_{i_2}(2+\frac{\lambda}{2})_{i_1}(\frac{7}{4}+\frac{\lambda}{2})_{i_1}}{\left(-\frac{\alpha }{4} +1+\frac{\lambda }{2}\right)_{i_1} \left( \frac{\alpha }{4} +\frac{5}{4}+\frac{\lambda }{2}\right)_{i_1}(2+\frac{\lambda}{2})_{i_2}(\frac{7}{4}+\frac{\lambda}{2})_{i_2}} \nonumber\\
&&\times \left. \sum_{i_3=i_2}^{\infty}  \frac{\left(-\frac{\alpha }{4} +\frac{3}{2}+\frac{\lambda }{2}\right)_{i_3} \left( \frac{\alpha }{4} +\frac{7}{4}+\frac{\lambda }{2}\right)_{i_3}(\frac{5}{2}+\frac{\lambda}{2})_{i_2}(\frac{9}{4}+\frac{\lambda}{2})_{i_2}}{\left(-\frac{\alpha }{4} +\frac{3}{2}+\frac{\lambda }{2}\right)_{i_2}  \left( \frac{\alpha }{4} +\frac{7}{4}+\frac{\lambda }{2}\right)_{i_2}(\frac{5}{2}+\frac{\lambda}{2})_{i_3}(\frac{9}{4}+\frac{\lambda}{2})_{i_3}}\eta ^{i_3} \right\} \mu ^3 \label{er:68d}
\end{eqnarray}
\end{subequations}
Put $l=1$ in (\ref{er:67}). Take the new (\ref{er:67}) into (\ref{er:68b}).
\begin{eqnarray}
 y_1(z)&=& c_0 z^{\lambda }\int_{0}^{1} dt_1\;t_1^{-\frac{3}{4}+\frac{\lambda }{2}} \int_{0}^{1} du_1\;u_1^{-\frac{1}{2}+\frac{\lambda }{2}} \nonumber\\
&&\times \frac{1}{2\pi i} \oint dv_1\;\frac{1}{v_1} \left(\frac{v_1-1}{v_1}\right)^{\frac{1}{2}(\frac{\alpha }{2}-1-\lambda )} (1-\eta v_1 (1-t_1)(1-u_1))^{-\frac{1}{2}(\frac{\alpha }{2}+\frac{3}{2} +\lambda )} \nonumber\\
&&\times  \left\{ \sum_{i_0=0}^{\infty }\Bigg( \Big(i_0+\frac{\lambda }{2}\Big)^2 -\Gamma ^{(I)} \Bigg) \right. \left. \frac{\left(-\frac{\alpha }{4} +\frac{\lambda }{2}\right)_{i_0} \left( \frac{\alpha }{4}+\frac{1}{4}+\frac{\lambda }{2}\right)_{i_0}}{(1+\frac{\lambda }{2})_{i_0}(\frac{3}{4} +\frac{\lambda }{2})_{i_0}} \left(\frac{t_1 u_1 v_1}{(v_1-1)} \frac{\eta }{1-\eta v_1 (1-t_1)(1-u_1)}\right)^{i_0} \right\} \mu \nonumber\\
&=& c_0 z^{\lambda } \int_{0}^{1} dt_1\;t_1^{-\frac{3}{4}+\frac{\lambda }{2}} \int_{0}^{1} du_1\;u_1^{-\frac{1}{2}+\frac{\lambda }{2}} \nonumber\\
&&\times \frac{1}{2\pi i} \oint dv_1\;\frac{1}{v_1} \left(\frac{v_1-1}{v_1}\right)^{\frac{1}{2}(\frac{\alpha }{2}-1-\lambda )} (1-\eta v_1 (1-t_1)(1-u_1))^{-\frac{1}{2}(\frac{\alpha }{2}+\frac{3}{2} +\lambda )} \nonumber\\
&&\times  \Bigg( \overleftrightarrow {w}_{1,1}^{-\frac{\lambda }{2}}\left(  \overleftrightarrow {w}_{1,1} \partial _{ \overleftrightarrow {w}_{1,1}}\right)^2 \overleftrightarrow {w}_{1,1}^{\frac{\lambda }{2}}-\Gamma ^{(I)}\Bigg) \left\{ \sum_{i_0=0}^{\infty } \frac{\left(-\frac{\alpha }{4} +\frac{\lambda }{2}\right)_{i_0} \left( \frac{\alpha }{4}+\frac{1}{4}+\frac{\lambda }{2}\right)_{i_0}}{(1+\frac{\lambda }{2})_{i_0}(\frac{3}{4} +\frac{\lambda }{2})_{i_0}} \overleftrightarrow {w}_{1,1} ^{i_0} \right\} \mu \label{er:69}
\end{eqnarray}
where 
\begin{equation}
 \overleftrightarrow {w}_{1,1} = \frac{t_1 u_1 v_1}{(v_1-1)}\; \frac{\eta }{1-\eta v_1 (1-t_1)(1-u_1)} \nonumber
\end{equation}
Put $l=2$ in (\ref{er:67}). Take the new (\ref{er:67}) into (\ref{er:68c}). 
\begin{eqnarray}
y_2(z) &=& c_0 z^{\lambda } \int_{0}^{1} dt_2\;t_2^{-\frac{1}{4}+\frac{\lambda }{2}} \int_{0}^{1} du_2\;u_2^{\frac{\lambda }{2}} \nonumber\\
&&\times \frac{1}{2\pi i} \oint dv_2\;\frac{1}{v_2} \left(\frac{v_2-1}{v_2}\right)^{\frac{1}{2}(\frac{\alpha }{2}-2-\lambda )} (1-\eta v_2 (1-t_2)(1-u_2))^{-\frac{1}{2}(\frac{\alpha }{2}+\frac{5}{2}+\lambda )} \nonumber\\
&&\times  \Bigg( \overleftrightarrow {w}_{2,2}^{-\frac{1}{2}(1+\lambda )}\left(  \overleftrightarrow {w}_{2,2} \partial _{ \overleftrightarrow {w}_{2,2}}\right)^2 \overleftrightarrow {w}_{2,2}^{\frac{1}{2}(1+\lambda )} -\Gamma ^{(I)}\Bigg)\nonumber\\
&&\times \left\{ \sum_{i_0=0}^{\infty} \frac{ (i_0+\frac{\lambda }{2})^2- \Gamma ^{(I)}}{(i_0+\frac{1}{2}+\frac{\lambda }{2})(i_0+\frac{1}{4}+\frac{\lambda }{2})}\frac{\left(-\frac{\alpha }{4} +\frac{\lambda }{2}\right)_{i_0} \left( \frac{\alpha }{4}+\frac{1}{4}+\frac{\lambda }{2}\right)_{i_0}}{(1+\frac{\lambda }{2})_{i_0}(\frac{3}{4} +\frac{\lambda }{2})_{i_0}}\right. \nonumber\\
&&\times \left.\sum_{i_1=i_0}^{\infty} \frac{\left(-\frac{\alpha }{4} +\frac{1}{2}+\frac{\lambda }{2}\right)_{i_1} \left( \frac{\alpha }{4} +\frac{3}{4}+\frac{\lambda }{2}\right)_{i_1}(\frac{3}{2}+\frac{\lambda}{2})_{i_0}(\frac{5}{4}+\frac{\lambda}{2})_{i_0}}{\left(-\frac{\alpha }{4} +\frac{1}{2}+\frac{\lambda }{2}\right)_{i_0} \left( \frac{\alpha }{4} +\frac{3}{4}+\frac{\lambda }{2}\right)_{i_0}(\frac{3}{2}+\frac{\lambda}{2})_{i_1}(\frac{5}{4}+\frac{\lambda}{2})_{i_1}} \overleftrightarrow {w}_{2,2}^{i_1} \right\} \mu^2 \label{er:70}
\end{eqnarray}
where
\begin{equation}
 \overleftrightarrow {w}_{2,2} = \frac{t_2 u_2 v_2}{(v_2-1)} \;\frac{\eta }{1-\eta v_2 (1-t_2)(1-u_2)}\nonumber
\end{equation} 
Put $l=1$ and $\eta =\overleftrightarrow {w}_{2,2}$ in (\ref{er:67}). Take the new (\ref{er:67}) into (\ref{er:70}).
\begin{eqnarray}
y_2(z) &=& c_0 z^{\lambda } \int_{0}^{1} dt_2\;t_2^{-\frac{1}{4}+\frac{\lambda }{2}} \int_{0}^{1} du_2\;u_2^{\frac{\lambda }{2}} \nonumber\\
&&\times \frac{1}{2\pi i} \oint dv_2\;\frac{1}{v_2} \left(\frac{v_2-1}{v_2}\right)^{\frac{1}{2}(\frac{\alpha }{2}-2-\lambda )} (1-\eta v_2 (1-t_2)(1-u_2))^{-\frac{1}{2}(\frac{\alpha }{2} +\frac{5}{2}+\lambda )} \nonumber\\
&&\times  \Bigg( \overleftrightarrow {w}_{2,2}^{-\frac{1}{2}(1+\lambda )}\left(  \overleftrightarrow {w}_{2,2} \partial _{ \overleftrightarrow {w}_{2,2}}\right)^2 \overleftrightarrow {w}_{2,2}^{\frac{1}{2}(1+\lambda )} -\Gamma ^{(I)}\Bigg)\nonumber\\ 
&&\times \int_{0}^{1} dt_1\;t_1^{-\frac{3}{4}+\frac{\lambda }{2}} \int_{0}^{1} du_1\;u_1^{-\frac{1}{2}+\frac{\lambda }{2}}\nonumber\\
&&\times \frac{1}{2\pi i} \oint dv_1\;\frac{1}{v_1}\left(\frac{v_1-1}{v_1}\right)^{\frac{1}{2}(\frac{\alpha }{2}-1-\lambda )} (1-\overleftrightarrow {w}_{2,2} v_1 (1-t_1)(1-u_1))^{-\frac{1}{2}(\frac{\alpha }{2}+\frac{3}{2} +\lambda )} \nonumber\\
&&\times \Bigg( \overleftrightarrow {w}_{1,2}^{-\frac{\lambda }{2}}\left(  \overleftrightarrow {w}_{1,2} \partial _{ \overleftrightarrow {w}_{1,2}}\right)^2 \overleftrightarrow {w}_{1,2}^{\frac{\lambda }{2}} -\Gamma ^{(I)}\Bigg) \nonumber\\
&&\times \left\{ \sum_{i_0=0}^{\infty } \frac{\left(-\frac{\alpha }{4} +\frac{\lambda }{2}\right)_{i_0} \left( \frac{\alpha }{4}+\frac{1}{4}+\frac{\lambda }{2}\right)_{i_0}}{(1+\frac{\lambda }{2})_{i_0}(\frac{3}{4} +\frac{\lambda }{2})_{i_0}} \overleftrightarrow {w}_{1,2} ^{i_0} \right\} \mu^2 \label{er:71}
\end{eqnarray}
where
\begin{equation}
 \overleftrightarrow {w}_{1,2}=\frac{t_1 u_1 v_1}{(v_1-1)}\; \frac{\overleftrightarrow {w}_{2,2}}{1-\overleftrightarrow {w}_{2,2} v_1 (1-t_1)(1-u_1)}\nonumber
\end{equation} 
By using similar process for the previous cases of integral forms of $y_1(z)$ and $y_2(z)$, the integral form of sub-power series expansion of $y_3(z)$ is
\begin{eqnarray}
y_3(z)&=& c_0 z^{\lambda } \int_{0}^{1} dt_3\;t_3^{\frac{1}{4}+\frac{\lambda }{2}} \int_{0}^{1} du_3\;u_3^{\frac{1}{2}+\frac{\lambda }{2}}\nonumber\\
&&\times \frac{1}{2\pi i} \oint dv_3\;\frac{1}{v_3}  \left(\frac{v_3-1}{v_3}\right)^{\frac{1}{2}(\frac{\alpha }{2}-3-\lambda )} (1-\eta v_3 (1-t_3)(1-u_3))^{-\frac{1}{2}(\frac{\alpha }{2}+\frac{7}{2} +\lambda )}\nonumber\\
&&\times \Bigg( \overleftrightarrow {w}_{3,3}^{-\frac{1}{2}(2+\lambda )}\left(  \overleftrightarrow {w}_{3,3} \partial _{ \overleftrightarrow {w}_{3,3}}\right)^2 \overleftrightarrow {w}_{3,3}^{\frac{1}{2}(2+\lambda )} -\Gamma ^{(I)}\Bigg)\nonumber\\
&&\times  \int_{0}^{1} dt_2\;t_2^{-\frac{1}{4}+\frac{\lambda }{2}} \int_{0}^{1} du_2\;u_2^{\frac{\lambda }{2}}  \nonumber\\
&&\times\frac{1}{2\pi i} \oint dv_2\;\frac{1}{v_2}  \left(\frac{v_2-1}{v_2}\right)^{\frac{1}{2}(\frac{\alpha }{2}-2-\lambda )} (1-\overleftrightarrow {w}_{3,3} v_2 (1-t_2)(1-u_2))^{-\frac{1}{2}(\frac{\alpha }{2}+\frac{5}{2} +\lambda )} \nonumber\\
&&\times \Bigg( \overleftrightarrow {w}_{2,3}^{-\frac{1}{2}(1+\lambda )}\left(  \overleftrightarrow {w}_{2,3} \partial _{ \overleftrightarrow {w}_{2,3}}\right)^2 \overleftrightarrow {w}_{2,3}^{\frac{1}{2}(1+\lambda )} -\Gamma ^{(I)}\Bigg)\nonumber\\
&&\times \int_{0}^{1} dt_1\;t_1^{-\frac{3}{4}+\frac{\lambda }{2}} \int_{0}^{1} du_1\;u_1^{-\frac{1}{2}+\frac{\lambda }{2}} \nonumber\\
&&\times \frac{1}{2\pi i} \oint dv_1\;\frac{1}{v_1} \left(\frac{v_1-1}{v_1}\right)^{\frac{1}{2}(\frac{\alpha }{2}-1-\lambda )} (1-\overleftrightarrow {w}_{2,3} v_1 (1-t_1)(1-u_1))^{-\frac{1}{2}(\frac{\alpha }{2}+\frac{3}{2} +\lambda )} \nonumber\\
&&\times \Bigg( \overleftrightarrow {w}_{1,3}^{-\frac{\lambda }{2}}\left(  \overleftrightarrow {w}_{1,3} \partial _{ \overleftrightarrow {w}_{1,3}}\right)^2 \overleftrightarrow {w}_{1,3}^{\frac{\lambda }{2}} -\Gamma ^{(I)}\Bigg) \nonumber\\
&&\times  \left\{ \sum_{i_0=0}^{\infty } \frac{\left(-\frac{\alpha }{4} +\frac{\lambda }{2}\right)_{i_0} \left( \frac{\alpha }{4}+\frac{1}{4}+\frac{\lambda }{2}\right)_{i_0}}{(1+\frac{\lambda }{2})_{i_0}(\frac{3}{4} +\frac{\lambda }{2})_{i_0}} \overleftrightarrow{w}_{1,3}^{i_0} \right\} \mu^3 \label{er:72}
\end{eqnarray}
where
\begin{equation}
\begin{cases} \overleftrightarrow {w}_{3,3} = \frac{t_3 u_3 v_3}{(v_3-1)}\; \frac{\eta }{1- \eta v_3 (1-t_3)(1-u_3)}  \cr
\overleftrightarrow {w}_{2,3} = \frac{t_2 u_2 v_2}{(v_2-1)}\; \frac{\overleftrightarrow{w}_{3,3} }{1- \overleftrightarrow{w}_{3,3} v_2 (1-t_2)(1-u_2)} \cr
\overleftrightarrow w_{1,3}= \frac{t_1 u_1 v_1}{(v_1-1)}\frac{\overleftrightarrow w_{2,3} }{1-\overleftrightarrow w_{2,3} v_1 (1-t_1)(1-u_1)}
\end{cases}
\nonumber
\end{equation}
By repeating this process for all higher terms of integral forms of sub-summation $y_m(z)$ terms where $m \geq 4$, we obtain every integral forms of $y_m(z)$ terms. 
Since we substitute (\ref{er:68a}), (\ref{er:69}), (\ref{er:71}), (\ref{er:72}) and including all integral forms of $y_m(x)$ terms where $m \geq 4$ into (\ref{eq:13}), we obtain (\ref{eq:73}).\footnote{Or replace the finite summation with an interval $[0,\alpha _0]$ by infinite summation with an interval  $[0,\infty ]$ in (\ref{eq:69}). Replace $\alpha _0$, $\alpha _{n-k}$ and $\alpha _{n-k-1}$ by $\displaystyle{\frac{1}{2}\Big(\frac{\alpha }{2}-\lambda \Big)}$, $\displaystyle{\frac{1}{2}\Big(\frac{\alpha }{2}-n+k-\lambda \Big)}$ and $\displaystyle{\frac{1}{2}\Big(\frac{\alpha }{2}-n+k+1-\lambda \Big)}$ into the new (\ref{eq:69}). Its solution is also equivalent to (\ref{eq:73}).}
\qed
\end{pot} 
Put $c_0$= 1 as $\lambda =0$  for the first independent solution of Lame equation and $\lambda =\frac{1}{2}$ for the second one into (\ref{eq:73}). 
\begin{rmk}
The integral representation of the first kind of Lame equation in the algebraic form about $x=a$ for the infinite series is
\begin{eqnarray}
 y(z)&=& LF \left( a, b, c, q, \alpha,\Gamma ^{(I)} = \frac{a}{2^4(2a-b-c)}\left( \alpha (\alpha +1) -\frac{q}{a}\right) ; z=x-a, \mu = \frac{-(2a-b-c)z}{(a-b)(a-c)}, \eta = \frac{-z^2}{(a-b)(a-c)} \right) \nonumber\\
&=& _2F_1 \left( -\frac{\alpha }{4}, \frac{\alpha }{4}+\frac{1}{4};\frac{3}{4}; \eta \right) + \sum_{n=1}^{\infty } \Bigg\{\prod _{k=0}^{n-1} \Bigg\{ \int_{0}^{1} dt_{n-k}\;t_{n-k}^{\frac{1}{2}(n-k-\frac{5}{2})} \int_{0}^{1} du_{n-k}\;u_{n-k}^{\frac{1}{2}(n-k-2)} \nonumber\\
&&\times  \frac{1}{2\pi i}  \oint dv_{n-k} \frac{1}{v_{n-k}} \left(\frac{ v_{n-k}-1 }{v_{n-k}} \right)^{\frac{1}{2}(\frac{\alpha }{2}-n+k)}    \left( 1-\overleftrightarrow {w}_{n-k+1,n} v_{n-k}(1-t_{n-k})(1-u_{n-k})\right)^{-\frac{1}{2}(\frac{\alpha }{2}+\frac{1}{2}+n-k)} \nonumber\\
&&\times \Bigg( \overleftrightarrow {w}_{n-k,n}^{-\frac{1}{2}(n-k-1)}\left(  \overleftrightarrow {w}_{n-k,n} \partial _{ \overleftrightarrow {w}_{n-k,n}}\right)^2 \overleftrightarrow {w}_{n-k,n}^{\frac{1}{2}(n-k-1)}-\Gamma ^{(I)}\Bigg) \Bigg\} \; _2F_1 \left( -\frac{\alpha }{4}, \frac{\alpha }{4}+\frac{1}{4};\frac{3}{4}; \overleftrightarrow {w}_{1,n}\right) \Bigg\} \mu ^n \nonumber 
\end{eqnarray}
\end{rmk}
\begin{rmk}
The integral representation of the second kind of Lame equation in the algebraic form about $x=a$ for the infinite series is
\begin{eqnarray}
y(z)&=& LS \left( a, b, c, q, \alpha,\Gamma ^{(I)} = \frac{a}{2^4(2a-b-c)}\left( \alpha (\alpha +1) -\frac{q}{a}\right) ; z=x-a, \mu = \frac{-(2a-b-c)z}{(a-b)(a-c)}, \eta = \frac{-z^2}{(a-b)(a-c)} \right) \nonumber\\
&=& z^{\frac{1}{2}} \Bigg\{\; _2F_1 \left(-\frac{\alpha }{4}+\frac{1}{4}, \frac{\alpha }{4}+\frac{1}{2};\frac{5}{4}; \eta \right) 
+  \sum_{n=1}^{\infty } \Bigg\{\prod _{k=0}^{n-1} \Bigg\{ \int_{0}^{1} dt_{n-k}\;t_{n-k}^{\frac{1}{2}(n-k-2)} \int_{0}^{1} du_{n-k}\;u_{n-k}^{\frac{1}{2}(n-k-\frac{3}{2} )} \nonumber\\
&&\times  \frac{1}{2\pi i}  \oint dv_{n-k} \frac{1}{v_{n-k}} \left(\frac{ v_{n-k}-1 }{v_{n-k}} \right)^{\frac{1}{2}(\frac{\alpha }{2}-\frac{1}{2}-n+k)} \left( 1-\overleftrightarrow {w}_{n-k+1,n} v_{n-k}(1-t_{n-k})(1-u_{n-k})\right)^{- \frac{1}{2}(\frac{\alpha }{2}+1+n-k)}\nonumber\\
&&\times \nonumber\\
&&\times \Bigg( \overleftrightarrow {w}_{n-k,n}^{-\frac{1}{2}(n-k-\frac{1}{2})}\left(  \overleftrightarrow {w}_{n-k,n} \partial _{ \overleftrightarrow {w}_{n-k,n}}\right)^2 \overleftrightarrow {w}_{n-k,n}^{\frac{1}{2}(n-k-\frac{1}{2})}-\Gamma ^{(I)}\Bigg) \Bigg\}  \; _2F_1 \left(-\frac{\alpha }{4}+\frac{1}{4}, \frac{\alpha }{4}+\frac{1}{2};\frac{5}{4}; \overleftrightarrow {w}_{1,n}\right) \Bigg\} \mu ^n \Bigg\}\nonumber
\end{eqnarray}
\end{rmk}
\section{Application to Laplace equation in ellipsoidal coordinates and nonlinear evolution equations}
A great many authors have worked on applications to boundary value problems for the Laplace equation in elliptic coordinates. We can apply the power series expansion in closed forms of Lame function and its integral forms into many mathematical physics areas. For example, Lame equation can be employed to solve boundary-value problems for Laplace equation in elliptical cones. 
In ``Occurrence of periodic Lame functions at bifurcations in chaotic Hamiltonian systems''\cite{Brac2001}, the authors shows that Lame equation in Weierstrass's form occurs at bifurcations in chaotic two-dimensional Hamiltonian systems with mixed phase space. (see (6), (7) in Ref.\cite{Brac2001}); Lame equation in the algebraic form can be transforms to in Weierstrass's form by changing an independent variable. Then we can describe Lame function more analytically in power series expansions in closed forms and its integral forms. 
In ``Lame Function and Its Application to Some Nonlinear Evolution Equations''\cite{Qian2003}, the Lame equation is applied to solve nonlinear $(1+1)$-dimensional, $(1+2)$- dimensional and coupled evolution equations. (see (1), (3), (5), (6) in Ref.\cite{Qian2003})

\section{Conclusion}
The power series expansion of Lame equation (ellipsoidal harmonics equation) has a recursive relation between a 3-term as we see (\ref{eq:3}). Because of its three term recurrence, there is a mathematical difficulty that Lame functions are described in the power series expansion in closed forms and its integral forms. 

In this paper I derive the power series expansion in closed forms for the infinite series and the polynomial of Lame function analytically by applying three term recurrence formula\cite{chou2012b}. 
Even if coefficient $\alpha $ can have two different values in which are $2 (2\alpha _i+i +\lambda )$ or  $-2 (2\alpha _i+i +\lambda )-1$ where $i,\alpha _i= 0,1,2,\cdots$to make $B_n$ term terminated at certain eigenvalues, its solutions are equivalent to each other as we see (\ref{eq:7}). 

Also as we see representations in the form of integrals in Lame functions, a $_2F_1$ function recurs in each of sub-integral forms: the first sub-integral form contains zero term of $A_n's$, the second one contains one term of $A_n$'s, the third one contains two terms of $A_n$'s, etc. Then we can transform Lame functions to all other well-know special functions such as Hypergeometric, Bessel, Legendre, Kummer functions, etc. 

Since we get the integral forms of power series expansions in Lame function, we are able to obtain generating functions of it. The generating functions are really helpful in order to derive orthogonal relations, recursion relations and expectation values of physical quantities.

Most of well-known papers with boundary value problems in ellipsoidal geometry have been published with using Weierstrass's form of it. So we need know what the power series expansion in closed forms of Lame function in Weierstrass's form and its integral forms. 
In Ref.\cite{Chou2012g} I derive the power series expansion in closed forms of Lame function in Weierstrass's form and its integral forms analytically. In Ref.\cite{Chou2012h} I construct the generating functions of Lame polynomial which makes $B_n$ term terminated in Weierstrass's form.

\section{Series ``Special functions and three term recurrence formula (3TRF)''} 

This paper is 6th out of 10.
\vspace{3mm}

1. ``Approximative solution of the spin free Hamiltonian involving only scalar potential for the $q-\bar{q}$ system'' \cite{chou2012a} - In order to solve the spin-free Hamiltonian with light quark masses we are led to develop a totally new kind of special function theory in mathematics that generalize all existing theories of confluent hypergeometric types. We call it the Grand Confluent Hypergeometric Function. Our new solution produces previously unknown extra hidden quantum numbers relevant for description of supersymmetry and for generating new mass formulas.
\vspace{3mm}

2. ``Generalization of the three-term recurrence formula and its applications'' \cite{chou2012b} - Generalize three term recurrence formula in linear differential equation.  Obtain the exact solution of the three term recurrence for polynomials and infinite series.
\vspace{3mm}

3. ``The analytic solution for the power series expansion of Heun function'' \cite{chou2012c} -  Apply three term recurrence formula to the power series expansion in closed forms of Heun function (infinite series and polynomials) including all higher terms of $A_n$'s.
\vspace{3mm}

4. ``Asymptotic behavior of Heun function and its integral formalism'', \cite{Chou2012d} - Apply three term recurrence formula, derive the integral formalism, and analyze the asymptotic behavior of Heun function (including all higher terms of $A_n$'s). 
\vspace{3mm}

5. ``The power series expansion of Mathieu function and its integral formalism'', \cite{Chou2012e} - Apply three term recurrence formula, analyze the power series expansion of Mathieu function and its integral forms.  
\vspace{3mm}

6. ``Lame equation in the algebraic form'' \cite{Chou2012f} - Applying three term recurrence formula, analyze the power series expansion of Lame function in the algebraic form and its integral forms.
\vspace{3mm}

7. ``Power series and integral forms of Lame equation in Weierstrass's form and its asymptotic behaviors'' \cite{Chou2012g} - Applying three term recurrence formula, derive the power series expansion of Lame function in Weierstrass's form and its integral forms. 
\vspace{3mm}

8. ``The generating functions of Lame equation in Weierstrass's form'' \cite{Chou2012h} - Derive the generating functions of Lame function in Weierstrass's form (including all higher terms of $A_n$'s).  Apply integral forms of Lame functions in Weierstrass's form.
\vspace{3mm}

9. ``Analytic solution for grand confluent hypergeometric function'' \cite{Chou2012i} - Apply three term recurrence formula, and formulate the exact analytic solution of grand confluent hypergeometric function (including all higher terms of $A_n$'s). Replacing $\mu $ and $\varepsilon \omega $ by 1 and $-q$, transforms the grand confluent hypergeometric function into Biconfluent Heun function.
\vspace{3mm}

10. ``The integral formalism and the generating function of grand confluent hypergeometric function'' \cite{Chou2012j} - Apply three term recurrence formula, and construct an integral formalism and a generating function of grand confluent hypergeometric function (including all higher terms of $A_n$'s). 
\vspace{3mm}

\bibliographystyle{model1a-num-names}
\bibliography{<your-bib-database>}
 
\end{document}